\documentclass[a4paper]{jpconf}
\usepackage{graphicx}

\usepackage{iopams}  
\usepackage[usenames,dvipsnames]{color}   
\usepackage{ulem}

\usepackage[utf8]{inputenc}
\usepackage[T1]{fontenc}
\usepackage{verbatim}
\usepackage{graphicx}

\usepackage{amsmath}
\usepackage{amssymb}
\usepackage{amsfonts}

\newcommand{\kB}{\ensuremath{k_{\text{B}}}}		
\newcommand{\eps}[0]{\varepsilon}	

\renewcommand{\vec}{\mathbf}
\newcommand{\db}[2][]{\text{d}^{#1}#2}

\renewcommand{\vec}[1]{\mathbf{#1}}

\begin{document}

\title[Wake Structure in Streaming Complex Plasmas]{On the Wake Structure in Streaming Complex Plasmas}

\author{Patrick Ludwig$^1$, Wojciech J. Miloch$^{2,3}$, Hanno K\"ahlert$^1$, and Michael Bonitz$^1$}

\address{$^1$ Institut f\"ur Theoretische Physik und Astrophysik, Christian-Albrechts-Universit\"{a}t zu Kiel, 24098 Kiel, Germany}

\address{$^2$ Department of Physics, University of Oslo, Box 1048 Blindern, 0316 Oslo, Norway}
\address{$^3$ Department of Physics and Technology, University of Troms\o, 9037 Troms\o, Norway}

\ead{ludwig@theo-physik.uni-kiel.de}

\begin{abstract}
The theoretical description of complex (dusty) plasmas requires multiscale concepts that adequately incorporate the correlated interplay of streaming electrons and ions, neutrals, and dust grains. 
Knowing the effective  dust-dust interaction, the multiscale problem can be effectively reduced to a one-component plasma model of the dust subsystem.  
The goal of the present publication is a systematic evaluation of the electrostatic potential distribution around a dust grain in the presence of a streaming plasma environment by means of two complementary approaches: 
(i) a high precision computation of the dynamically screened Coulomb potential from the dynamic dielectric function, and
(ii) full 3D particle-in-cell simulations, which self-consistently include dynamical grain charging  and non-linear effects.
The applicability of these two approaches is addressed.
\end{abstract}

\section{Introduction}
Complex (dusty) plasmas have been proven to be an instructive reference system for strong coupling and correlation effects \cite{ishihara2007,bonitz2010,ludwig2010a}. Thereby, complex plasma research inherently connects key issues from several fields including low temperature physics, warm dense matter physics,  surface and solid-state physics, as well as material science \cite{goreePT,fortov2005,ShuklaRev,MorfillRev,springer2010,Chaudhuri2011}. 

The first principle description of complex plasmas is a theoretically challenging problem. 
A dusty plasma consists of electrons, positive ions, neutral atoms, and micrometer-sized dust grains, i.e.,
components with distinct mass asymmetry giving rise to dynamics on very different spatial and temporal scales.
This multi-scale behavior, combined with the \textit{non-ideal character} of a complex plasma, makes a full temporal resolution on all scales a numerically unfeasible task with current computer technology. However, there are two numerical approaches that can address the challenges associated with the complex plasma system: the One-Component Plasma (OCP) model and the Particle-In-Cell (PIC) method.

The OCP approach relies on the idea to reduce the quasi-neutral, multi-component complex plasma to the subsystem of dust grains which interact via a screened Coulomb potential. In particular, the statically screened (Yukawa-type) potential yields good agreement with the experiments for various specific setups, see for example Refs. \cite{bonitz2006,donko2008}. Also, this isotropic potential is well suited for fundamental analytical and numerical investigations on cooperative effects such as self-organized structure formation \cite{ludwig2010a,henning2006}, collective dynamical processes \cite{henning2008, ott2009, kaehlert2010}, spectral properties \cite{magneto2010,kaehlert2011}, or the melting behavior~\cite{ogawa2006,apolinario2007} in strongly correlated (screened) Coulomb systems.
The multi-component and non-equilibrium nature of a complex plasma requires, however, a careful analysis of plasma streaming and dynamical grain charging effects, which were experimentally shown to become of utmost importance at least near the plasma sheath, see e.g.~\cite{taka98,melzerPRL99,hebner,mebu2010,Carstensen,Saitou}.
In a plasma with ions streaming at a uniform velocity, the dust potential becomes (strongly) anisotropic and takes the form of an oscillatory wake structure in the downstream direction, which was subject to various analytical and numerical studies e.g.~\cite{vlad1995, melands1995,schw1996, ishihara, lapenta1999, lemons, winske,lampe2000,ivlev2004,schweig2005, hutch2006, miloch2007,komp208,lapenta2008, guidoPRE2008, miloch2010, hutch2011} \footnote{For a review of previous experimental and theoretical work on the structure of the dust (wake) potential, we refer the following in-depth references \cite{fortov2005,ShuklaRev,MorfillRev,mebu2010,chaudhuri2010,vlad11}. A recommended critcal discussion of previous theoretical work on the structure of plasma wakes can be found in \cite{hutch2011}.}. This wake can result in attractive, non-reciprocal forces between the equally charged dust grains and lead to remarkable structural and dynamical consequences~\cite{kroll2010,schella2011,killer,joyce2001,Hammerberg2, lampe2005, miloch2008, miloch2010a, ludwig2011}. Streaming effects can be incorporated in the OCP dynamical screening model by means of Linear Response (LR) theory, as presented in section~\ref{DSM}.

The second powerful numerical ansatz to tackle the multi-scale problem  of the different plasma components is the PIC approach, in which the trajectories of plasma particles are followed in electric fields self-consistently \cite{verb2005}. To obtain a self-consistent charge on the grains, the plasma dynamics need to be resolved  in the vicinity of the grain on spatial and temporal scales which are smaller than the Debye length and electron plasma period, respectively. 
However, due to the large mass of grains (as compared to electrons and ions) as well as the large distances between the grains, the study 
of the dynamics of several dust grains requires simulation times that are much longer than the dust charging (which is on the ion plasma period time scale). 
Thus, a self-consistent 3D PIC calculation, i.e., a dynamic evolution of the 3D system with more than one dust grain with a temporal resolution on the electron or ion plasma period time scale, required to ensure self-consistency, easily exceeds the capabilities of present high-performance computing systems~\cite{miloch2010a, matyash2010}.

Consequently, short-time, small-scale PIC simulations of dust charging or, alternatively, less demanding LR calculations of the dynamical plasma shielding have to be coupled with large-scale OCP simulations which incorporate the interaction between the grains on a more abstract level \cite{joyce2001,Hammerberg2, lampe2005}.
Such a coupled multi-scale numerical approach may ensure the description of the correlated system dynamics with proper charges on the grains as well as an accurate potential distribution in the vicinity of grains.
With the goal to form a method-spanning picture, the present publication comprises a systematic analysis of the electrostatic potential distribution around a dust grain in the presence of a streaming plasma environment by 
\begin{enumerate}
 \item a high precision computation of the dynamically screened Coulomb potential from the dynamic dielectric function, and 
 \item a critical assessment of these LR results, in particular in the view of non-linear effects and  dynamical grain charging processes by means of self-consistent 3D PIC simulations. 
\end{enumerate}
The discussion of the LR results, section \ref{DSM}, extends over a broad range of plasma streaming velocities, different electron-to-ion temperature ratios and includes  an evaluation of the influence of collisional damping. The corresponding 3D PIC simulations, section \ref{sec:PIC}, are performed for the collisionless limit and apply, therefore, to the low pressure limit.
Finally, it is vital to verify whether and to what degree the results obtained by a linear superposition of single grain potentials do coincide with the corresponding full 3D PIC simulations. Such a direct comparison is performed for the case of two grains in section \ref{sec:discussion}.



\subsection{Plasma Parameters}\label{plasmaparameters}
The plasma parameters are scaled to relevant base units, which are, in our case, 
the Debye length $\lambda_{D_\alpha}=\sqrt{\eps_0 \kB T_\alpha/(n_\alpha q_\alpha^2)}$ 
and the plasma period $\tau_\alpha=2\pi/\omega_{p_\alpha}$, with the corresponding plasma frequency
$\omega_{p_\alpha}=\sqrt{{n_\alpha q_\alpha^2}/({\eps_0 m_\alpha})}$, where $\alpha$ denotes electrons (e) or ions (i). $T_\alpha$ is the corresponding temperature, $n_\alpha$ density, $m_\alpha$ mass, $q_\alpha$ charge, $\kB$ the Boltzmann constant, and $\eps_0$ is the permittivity of vacuum.
The relative streaming velocity $\vec u_i$ of the ions with respect to the dust is characterized by the Mach number $M \equiv u_i/c_s$,
where the ion sound (Bohm) speed is given by $c_s\equiv\sqrt{{\kB T_e}/{m_i}}$. Furthermore, the thermal velocity is $v_{T_\alpha}=\sqrt{{k_B T_\alpha}/{m_\alpha}}$. 

The dynamically screened dust potential and in particular the characteristics of the wake, are studied with respect to three dimensionless plasma parameters:
\begin{itemize}
 \item the ion drift velocity, expressed in terms of the Mach number $M$,
 \item the ratio of electron-to-ion temperature $T_r\equiv T_e/T_i$,
 \item the ion-neutral scattering frequency (normalized to the ion plasma frequency) $\nu_{in}/\omega_{p_i}$.
\end{itemize}
While the temperature ratio $T_e/T_i$ controls the effect of Landau damping, the frequency ratio $\nu_{in}/\omega_{p_i}$ determines the effect of collisional damping and can be effectively controlled in experiments by changing the neutral gas pressure. 
A simple approximation for the pressure as function of the ion-neutral collision frequency and ion temperature is given by 
\begin{align}\label{pressure}
p=\frac{\nu_{in}/\omega_{p_i}\cdot e_0 \sqrt{k_B T_i\cdot n_e/\eps_0}}{\sigma_{in}} \quad, 
\end{align}
where $e_0=1.602\cdot10^{-19}~\mathrm{C}$ is the elementary charge, and $\sigma_{in}\simeq 5.0\cdot 10^{-19}~\mathrm{m^2}$ is the ion-neutral collision cross-section as defined in Ref. \cite{sigma_in}.\footnote{We should note that the ion-neutral scattering frequency $\nu_{in}$ and not the roughly approximated pressure $p$ is used as input parameter for the LR calculations. For the energy dependence of the ion-neutral cross-section in Argon see e.g.~\cite{cramers}.}

For the LR calculations, we consider Argon, which is a typical working gas in complex plasma experiments \cite{springer2010}. The ions are considered to be singly charged ($q_i=e_0$). Consequently, ion and electron densities are equal, and are set to $n_e=n_i=2.0\cdot10^{14}~\mathrm{m^{-3}}$ and are considered as spatially homogeneous distributions. The considered electron temperature is $T_e=2.585~\mathrm{eV}$ ($\simeq30000~\mathrm{K}$). The corresponding electron Debye length is  $\lambda_{D_e}=845~\mathrm{\mu m}$.
An atomic mass of Argon $m_n=m_i=6.634\cdot10^{-26}~\mathrm{kg}$ leads to the Bohm speed of $c_s=2500~\mathrm{m/s}$. For the considered set of parameters, the ion plasma frequency takes the value of $\omega_{p_i}=3.0\cdot10^{6}~\mathrm{Hz}$. 
For the PIC simulations, we simulate ions with a reduced mass $m_i/m_e=120$, and thus $\omega_{p_i}$ and $c_s$ are increased accordingly (see section \ref{pictheory} for details). The LR and PIC results presented in this paper do not involve any kind of free fit parameters.

\subsection{Grain charging and wake formation}
The dust grains are charged by plasma currents and other currents, such as induced by the photoelectric effect or secondary electron emission. For the charge equilibrium, the net current to the grain is zero, and such a grain is at \textit{floating} potential $\Phi_{fl}$ with respect to the surrounding plasma. 
If charging is only due to plasma currents (i.e., electron and ion fluxes to the surface), then in electropositive plasmas, the charge $Q_d$ on the grain and the floating potential will be negative, due to the high mobility of electrons. 
The floating potential on a spherical, finite-sized grain can be approximated with the Orbit-Motion-Limited (OML) approach for both stationary and flowing collisionless plasmas \cite{miloch2010a}. We note that in the presence of collisions, the OML theory can give incorrect results, as it has been noted by several authors \cite{bernstein,goree1992,zobnin2000}. The presence of trapped ions created by ion-neutral collisions can increase the ion current to the grain, and thus reduce the charge on the grain \cite{lampe2003,khrapak2005,hutch2007}.

\begin{figure}[t]\hspace{-0.01\textwidth}%
\begin{minipage}{1.0\textwidth}
\includegraphics[width=0.8\textwidth]{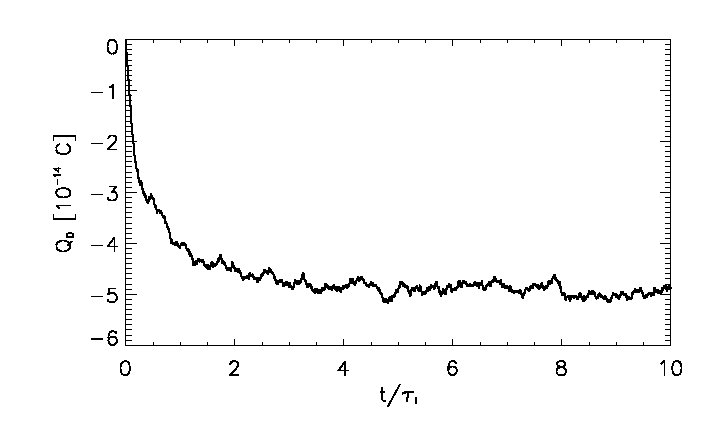}
\caption{\label{fig:chargingPIC} Typical charging characteristics for a spherical conducting grain in stationary plasma. Data is from 3D PIC simulations of a stationary plasma with $T_e/T_i=30$, time $t$ is normalized with the ion plasma period $\tau_i$.}
\end{minipage}\hspace{0.018\textwidth}%
\end{figure}

In a stationary plasma, the charging of a grain takes approximately one ion plasma period, $\tau_i$, a time in which the ions can collectively respond to the perturbations in the electric field. A typical charging characteristic for a single, spherical conducting grain is shown in Fig.~\ref{fig:chargingPIC}, where the data is taken from 3D PIC simulations for the stationary plasma with $T_e/T_i=30$.  
In the static case, the charge on a small grain can be calculated with the capacitance model for a given floating potential~\cite{bouchoule1999}:
\begin{equation}
Q_d=4\pi \epsilon_0 a \left( 1+ \frac{a}{\lambda_{{D}}}\right) \Phi_\mathrm{fl} \quad,
\end{equation}
where $a$ is the grain radius and $\lambda_{D}$ the total Debye length defined in terms of electron and ion Debye lengths $\lambda_{{D}}^{-2}  \equiv \lambda_{{D_e}}^{-2}+\lambda_{{D_i}}^{-2}$ for a static case, i.e., $M=0$. 
When charging is only due to plasma currents, the floating potential in a collisionless plasma is typically on the order of $\Phi_\mathrm{fl} \approx -2\kB T_e/e_0$. Thus, from the capacitance model one can derive a formula for an approximate charge on a negligible small grain with $a \ll   \lambda_{D}$:
\begin{equation}
Q_d/e_0=-1400 a_{[\mu m]} T_{e, [eV]} \quad,  
\end{equation}
where the grain radius $a$ is given in micrometers, and temperature in eV.
In the case of no plasma flow relative to the dust grain, the negatively charged grain is shielded by a stationary, positively charged ion cloud. The electric potential around a small (non-absorbing) charged grain takes the form of a statically screened Coulomb potential which is typically referred to as Debye-Hückel or Yukawa potential 
\begin{equation} \label{Yukawa}
\Phi_\mathrm{Yuk}(r)=\frac{1}{4\pi \eps_0} \frac{Q_d}{r} e^{-\frac{r}{\lambda_{D}}} \quad.
\end{equation}
Plasma absorption, ion-neutral collisions, and plasma flow can modify the potential distribution. In particular, the long-range asymptote of the plasma-mediated potential is often not screened exponentially, but has a power-law decay instead, see e.g.~\cite{lampe2000,chaudhuri2010,vlad11,joyce1968, khrapakPRL2008}.

The plasma flow breaks the symmetry in the grain charging. In the mezothermal velocity regime, when the flow velocity $u_i$ is larger than the ion thermal velocity $v_{{T_i}}$ and lower than the electron thermal velocity $v_{{T_e}}$ ($v_{{T_i}}< u_i < v_{{T_e}}$),  the grain will receive a larger ion flux to the surface on the upstream than on the downstream side, while the electron flux will be similar to both sides. Anisotropic charging can lead to an electric dipole on the grain \cite{lapenta1999, ivlev1999}. The electric dipole moment depends on the material of the grain and is more pronounced for supersonic flows \cite{miloch2007}. For small conducting grains the electric dipole can often be neglected, as the charge is redistributed on the surface, but for larger conducting grains, potential distribution in the vicinity of a grain can modify the charge distribution on the grain surface.

A streaming plasma leads to a wake in the density and potential distributions around the grain. The main mechanisms for the wake formation are plasma absorption on the grain surface and the influence of the electric field on plasma particle trajectories in the vicinity of the grain \cite{miloch2010}. Plasma absorption leads to density depletion on the downstream side, while electric fields can scatter ions into the wake region forming the ion focus region. Both effects are more pronounced for cold ions, as well as for supersonic flows, when also a Mach cone forms. 

The charging time in a streaming plasma is longer than in a stationary plasma and depends on the material of a dust grain, being longer for insulating grains, for which it can take several ion plasma periods to reach stationary conditions.


\section{Linear Response Approach} \label{linresponse}\label{DSM}
According to the OCP concept, the multi-component dusty plasma can be reduced to the subsystem of dust grains which interact via a dynamically screened Coulomb potential. This dynamical potential takes into account the effect of plasma streaming on the dielectric response of the plasma. 
Depending on the quality of dielectric function $\epsilon(\vec k,\omega)$, the considered dynamical screening model comprises an accurate representation of the most essential plasma properties including screening, wakefield oscillations, ion and electron thermal effects, Landau damping, as well as collisional damping \cite{lampe2000,lampe2005}.

\subsection{Theoretical Approach} \label{theoapproach}
\begin{figure}[t]\hspace{0\textwidth}%
\begin{minipage}{0.35\textwidth}
\includegraphics[width=1.25\textwidth]{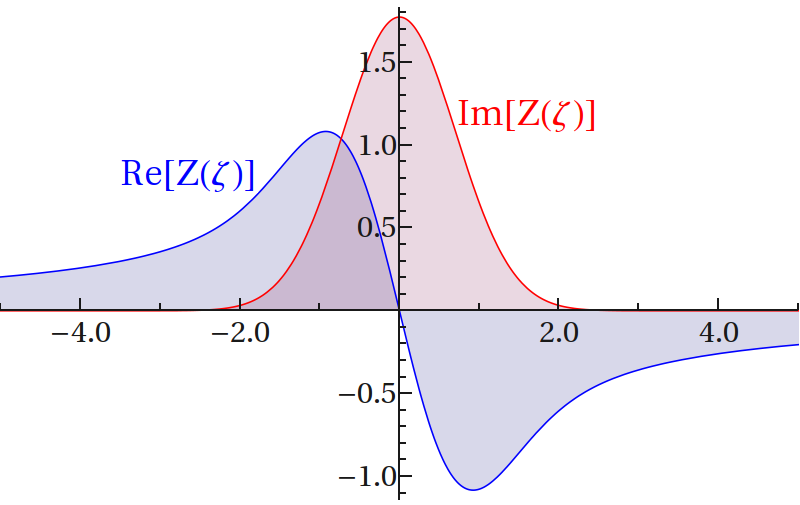}
\end{minipage}\hspace{0.12\textwidth}%
\begin{minipage}{0.35\textwidth}
\includegraphics[width=1.25\textwidth]{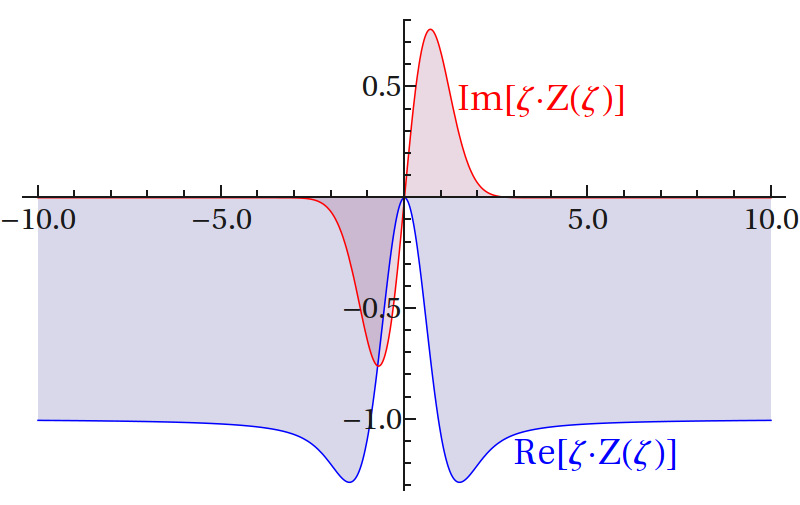}
\end{minipage}
\begin{minipage}[b]{\textwidth}\caption{\label{pdf}Real and imaginary part of the plasma dispersion function  $Z(\zeta)$ for a  real  $\zeta$  (left) and the product $\zeta \cdot Z(\zeta)$ (right) which enters Eq.~(\ref{dk}). 
}
\end{minipage}
\end{figure}

Considering a weak (linear) response of the plasma to the presence of the dust grain, the electric potential can be computed by a 3D Fourier transformation 
\begin{align} \label{potential}
\Phi(\vec r) = \int\!\db[3]{\vec{k}} \frac{Q_d}{2 \pi^2 k^2 \epsilon(\vec k, \vec k \cdot \vec u_i)} e^{i \vec k \vec r} \quad,
\end{align}
where $\vec r$ denotes the distance from the grain's center of mass. The (shifted) Maxwellian plasma is represented by the following longitudinal dielectric function which includes Bhatnagar–Gross–Krook (BGK)-type ion-neutral collisions~\cite{alexandrov,jenko2005}
\begin{align}\label{dk}
\epsilon^l(\vec{k},\omega)= 1 + \frac{1}{k^2 \lambda_{D_e}^2} + \frac{1}{k^2 \lambda_{D_i}^2} \left[ \frac{1+\zeta_i Z(\zeta_i)} {1+\frac{i\nu_{in}}{\sqrt{2} k v_{T_i}}Z(\zeta_i)} \right] \quad,
\end{align}
where we use the substitution
\begin{align}\label{zeta}
\zeta_i=\frac{\vec{k}(\vec{v}_d-\vec{u}_i)+i\nu_{in}}{\sqrt{2} k v_{T_i}} \quad.
\end{align}
The screening by electrons is considered as static (Yukawa-type), since the electron flow is of no relevance ($u_e \ll v_{T_e}$). The plasma dispersion function is defined as
\begin{align}
Z(\zeta)= \frac{1}{\sqrt{\pi}}\int_{-\infty}^\infty \frac{\exp(-t^2)}{t-\zeta} dt = i\sqrt{\pi} \exp{(-\zeta^2)} \text{Erfc}(-i \zeta) \quad.
\end{align}
The function $Z(\zeta)$ and the product $\zeta \cdot Z(\zeta)$ are plotted in Fig.~\ref{pdf}. In the limiting case of $M=0$, that is $(\vec{v}_d-\vec{u}_i) \rightarrow 0 $, Eq.~(\ref{dk}) reduces to static electron and ion shielding, which corresponds to the spherically symmetric Yukawa potential Eq.~(\ref{Yukawa}). For $|\vec{v}_d-\vec{u}_i| \gg v_{T_i}$ the product $\zeta_i\cdot Z(\zeta_i)$ approaches $-1$ such that the ion screening contribution vanishes and only electron screening remains in Eq.~(\ref{dk}), see right panel of Fig.~\ref{pdf}. The finite electron contribution ensures the numerical convergence of the Fourier integral Eq.~(\ref{potential}).

\subsection{Numerical implementation} \label{numimplr}
The accuracy of the Dynamical Screening Model depends crucially on the 3D Discrete Fourier Transform (DFT), Eq.~(\ref{potential}). A sufficient sampling rate in k- and r-space at the same time, requires a large number of sample points, which are, however, in 3D typically limited by computer resources to $N_{\rm x}=N_{\rm y}=N_{\rm z}=256$. Even with the optimal choice of the sampling interval, corresponding to the grid spacing in the reciprocal space, the result for many of the relevant plasma parameters is not satisfying and lacks accuracy. In particular, the cut off error in the streaming direction leads to pseudoperiodicity effects from the DFT, such as artificial oscillations in upstream direction.

A straightforward approach to reduce the numerical complexity of the Fourier integral Eq.~(\ref{potential}) is to make use of the axial symmetry in streaming direction by introducing cylindrical coordinates. However, the numerical evaluation of the resulting radial Hankel transform is very time-consuming and converges poorly due to an oscillating Bessel function in the integrand. Instead, an alternative approach was used here to improve the memory and time efficiency by exploiting the radial symmetry. 
The 3D integration area was sliced along $k_z$-direction and $N_{\rm z}$ 2D-DFTs were performed on a $k_{\rm x,y}=256 \times 256$ grid. Notably, for each $k_z$-value only the resulting $N_{\rm x}/2=128$ values in one radial direction (we used the positive half of the $x$-axis) have to be stored and used for the subsequent $k_z$-integration. Hence, instead of $N_{\rm x} \cdot N_{\rm y}=256^2$ only $N_{\rm x}/2=128$ DFTs in $k_z$-direction have to be computed. Moreover, using this scheme the integration area can be easily adjusted to asymmetric spatial dimensions. In particular, for $M\geq0.5$ in the case of weak (Landau and/or collisional) damping, we used $N_{\rm z}=512$ instead of $N_{\rm z}=256$ grid points which considerably reduced pseudoperiodicity effects and ensured convergence of $\Phi(\vec r)$. 

For the sake of completeness, we should mention that because of different length scales of the Debye potential and the wake modulations we did not transform $\Phi(\vec k)$ directly, but the difference $\Delta \Phi(\vec k) = \Phi(\vec k) - \Phi_{\rm Yuk}(\vec k) $ \cite{joyce2001}, where $\Phi_{\rm Yuk}(\vec k)$ denotes the collisionless static case. Subsequently, utilizing the linearity of the Fourier transform, the Yukawa potential Eq.~(\ref{Yukawa}) is added in real space again. In real space, the DFT yields a grid resolution of $\lambda_{D_e}/8$ for the wake contribution $\Delta \Phi(\vec r)$. The data is post processed with a spline interpolation.


In order to ensure the convergence of the  Fourier integration, sufficient collisional and/or Landau damping is required. 
More specifically, for the collisionless case ($\nu_{in} = 0$) the temperature ratio is limited to $T_e/T_i \leq 25$.  For reliable results over the full range of $M$ and temperature ratios up to $T_e/T_i = 100$, weak collisional damping $\nu_{in} = 0.1 \omega_{p_i}$ is required. The Fourier integration performs better for small values of $M$.

\subsection{Results}
The LR calculations discussed here use the plasma parameters (electron temperature, density, ion mass, etc.) as  introduced in section~\ref{plasmaparameters}. 
Neglecting the dust drift, i.e. $|\vec{v}_d|=0$, the only dust parameter which enters the calculations is the grain charge, which is set to $Q_d= -10^4 e_0=-1.602\cdot10^{-15}~\mathrm{C}$. Since the wake potential of a point-like grain ($a/\lambda_{D_e}\simeq0$) scales linearly with $Q_d$, a pre-computed potential $\Phi(\vec r)$ can easily be adapted to any other grain charge required, see Eq.~(\ref{potential}).

\begin{figure}[t]
\includegraphics[width=\textwidth]{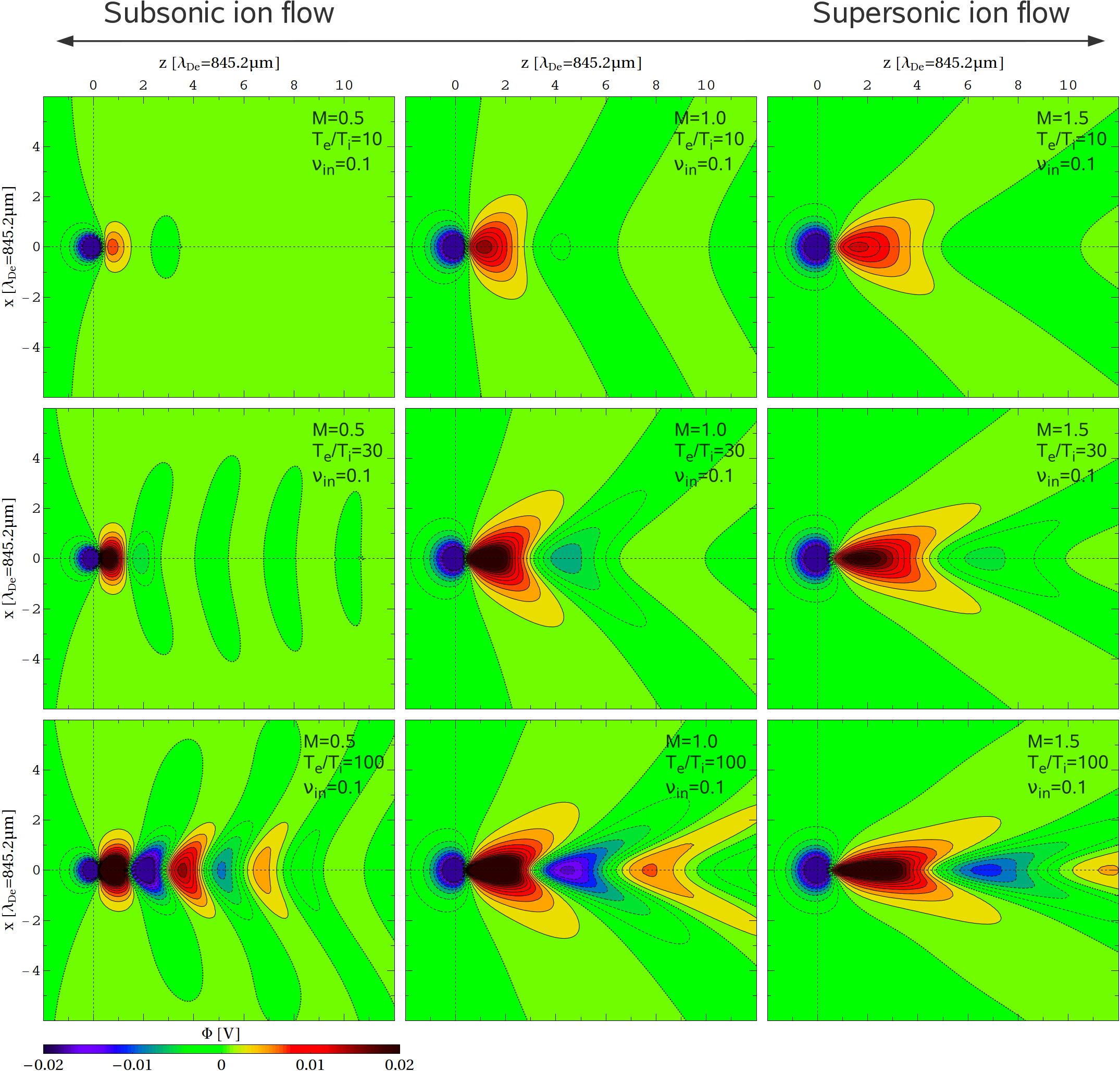}\hspace{2pc}%
\begin{minipage}[b]{\textwidth}\caption{\label{pic:matrix}Linear response results: contour plots of the grain potential $\Phi(\vec r)$ for varied streaming velocities $M=0.5, 1.0, 1.5$ (from left to right) and electron-ion temperature ratios $T_e/T_i=10, 30, 100$, that is $T_i=3000\,\mathrm{K}, 1000\,\mathrm{K}, 300\,\mathrm{K}$ (from top to bottom). The dust grain is placed at the origin and the plasma flows in the positive $z$ direction. The ion-neutral collision frequency is fixed to $\nu_{in}/\omega_{p_i}=0.1$. 
Dashed-blue (solid-red) contours denote areas of negative (positive) space charge; potential values larger than $20\,\mathrm{mV}$ are clipped and shown black. The dotted line (on green ground) corresponds to $\Phi=0$. The equipotential lines are separated by $2\,\mathrm{mV}$.}
\end{minipage}
\end{figure}

Fig.~\ref{pic:matrix} shows the shape of the dynamically screened (wake) potential  $\Phi(\vec r)$ in real space
for a range of typical plasma parameters taken from the experiments~\cite{springer2010,mebu2010}. To capture the general trends, we consider three characteristic electron-to-ion temperature ratios $T_r=T_e/T_i= 10, 30, 100$ (top to bottom row)  and drift velocities in the subsonic regime ($M=0.5$, left), at Bohm speed ($M=1.0$, middle column), and supersonic ion flow ($M=1.5$, right).
The ion-neutral scattering frequency is fixed at a value of $\nu_{in}/\omega_{p_i}=0.1$. According to Eq.~(\ref{pressure}) this corresponds to pressures of $\mathrm{p}=31\,\mathrm{Pa}, 18\,\mathrm{Pa}$ and $10\,\mathrm{Pa}$ for $T_{\rm r}=10, 30, 100$, respectively.
The plasma flow gives rise to an oscillating wake structure downstream the grain. On that account, the grain potential $\Phi(\vec r)$ essentially deviates from the static Yukawa potential in all considered cases. Typically, the potential's anisotropy, i.e. the symmetry breaking, increases with $M$. The wakefield is most pronounced for large values of $T_{\rm r}$.

For $M=0.5$ and $T_e/T_i=10$ ($\mathrm{p}=31\,\mathrm{Pa}$), the range and height of the wakefield oscillations are strongly reduced by the overlapping effect of Landau and collisional damping. Only a relatively weakly pronounced primary potential maximum, i.e. a single node directly behind the grain, emerges in the subsonic regime.\footnote{Compared to a isotropically screened Yukawa system, even a relatively shallow wake, such as observed for $M=0.5$ and $T_{\rm r}=10$, was recently shown to have a drastic effect on the collective many particle behavior~\cite{ludwig2011}. Note that for a subsystem dust grains, the non-reciprocality along the streaming direction causes a violation of Newton's law of that action equals reaction, see e.g.~\cite{lampe2005}.} For $M=1.0$, and even more pronounced for $M=1.5$, the wave front becomes cone-shaped. In the supersonic case, the peak is lower than for $M=1.0$. The same is observed for the second (third) potential peak for $T_r=30$ ($T_r=100$). 
Further maxima and minima besides the first downstream potential maximum start to appear for temperature ratios $T_{\rm r}>10$. 
Considering $T_r=100$, there are in total three (significant) positive space charge regions, which may result in an attractive (non-reciprocal) force between the equally charged dust grains. For $M=1.0$ the form of the wake extends far in the cross-streaming direction. With growing $M$ the wake becomes increasingly stretched in the streaming direction while the peak height goes down. A plane $\Phi=0$ wave front is observed for $M\approx0.875$ ($T_r=30$).

\begin{figure}[t]\hspace{-0.01\textwidth}%
\begin{minipage}{0.3\textwidth}
\includegraphics[width=1.25\textwidth]{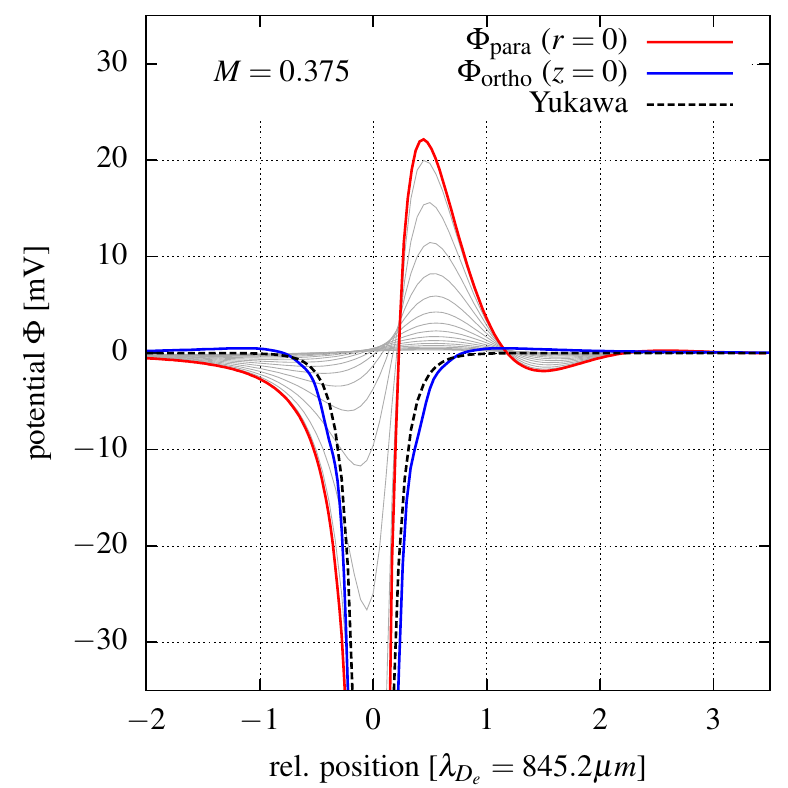}
\end{minipage}\hspace{0.018\textwidth}%
\begin{minipage}{0.3\textwidth}
\includegraphics[width=1.25\textwidth]{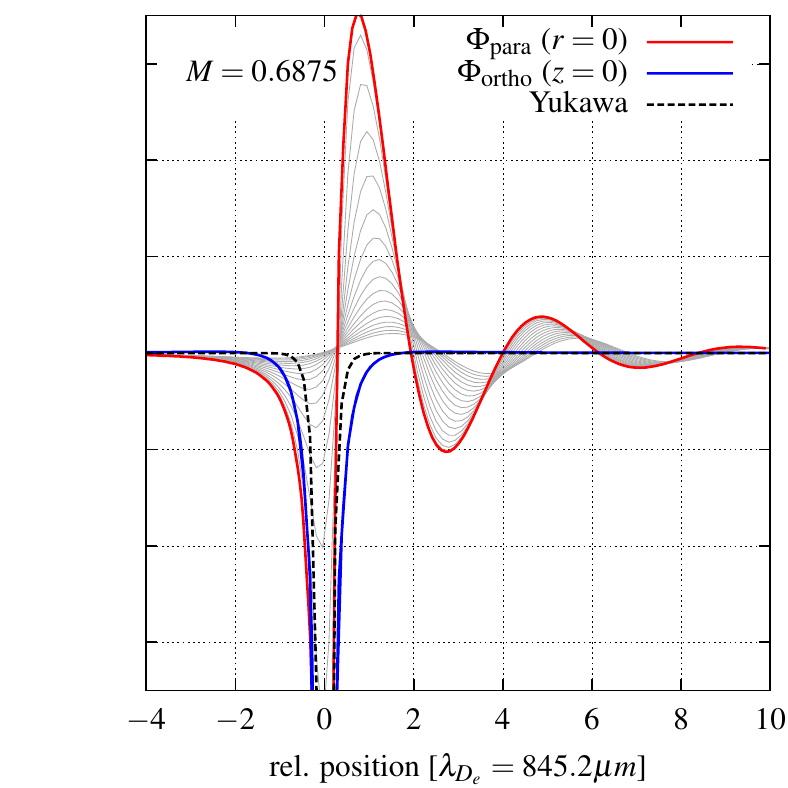}
\end{minipage}\hspace{0.018\textwidth}%
\begin{minipage}{0.3\textwidth}
\includegraphics[width=1.25\textwidth]{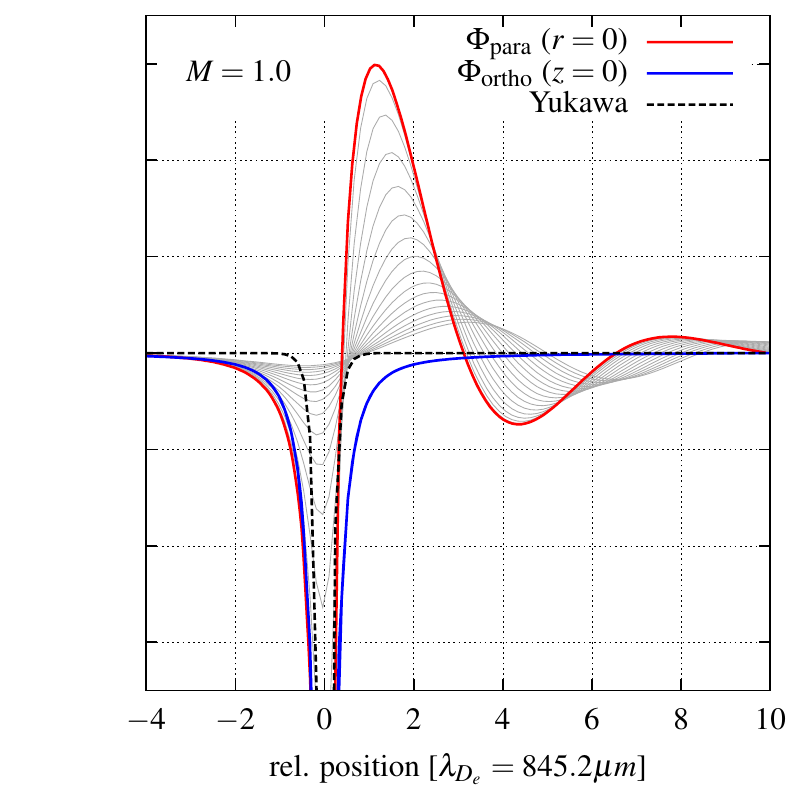}
\end{minipage}
\begin{minipage}[b]{\textwidth}\caption{\label{pic:sideview}LR results for the electrostatic grain potential in the streaming direction $z$ at $x = y =0$ (red) and perpendicular to the flow at $z=0$ (blue) for $M=0.375, 0.6875$, and $1.0$ (left to right panel). The dashed black lines indicate the (isotropic) Yukawa potential, Eq.~(\ref{Yukawa}), for the corresponding total Debye length $\lambda_{D}=0.8452\,\mathrm{mm}$. The light gray curves represent slices through the potential surface in $z$-direction for $y$-values in the range $2 \lambda_{D_e}\leq y<0$. Considered parameters: $T_r=30$ and $\nu_{in}=0.1\omega_{p_i}$ (cf. middle row in Fig.~\ref{pic:matrix}). 
Note the different scaling of the abscissas; the range of the ordinate is $[-35\,\mathrm{mV}:35\,\mathrm{mV}]$.
}
\end{minipage}
\end{figure}

\begin{figure}[t]
\begin{minipage}{0.3\textwidth}
\includegraphics[width=1.25\textwidth]{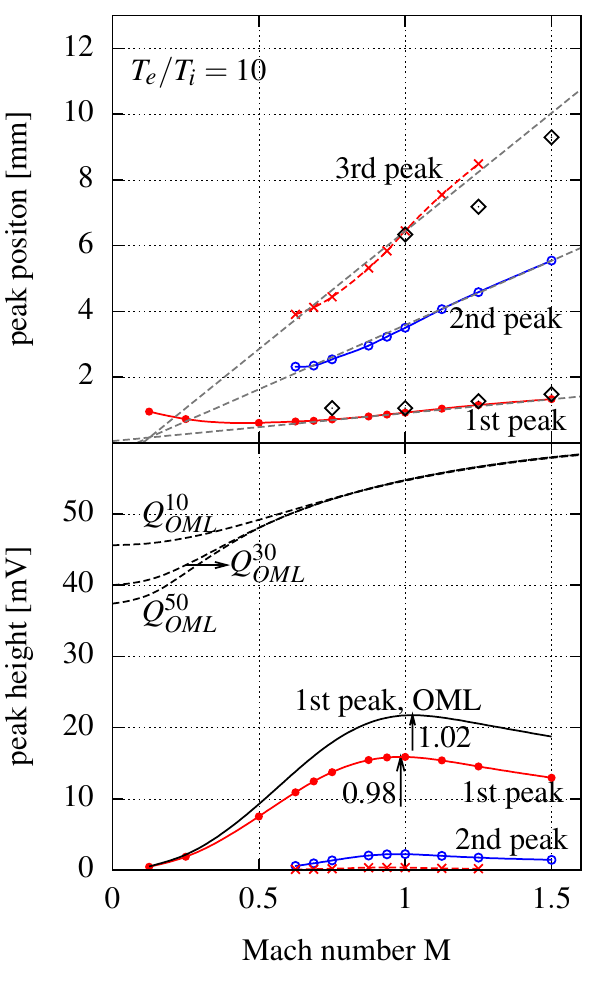}
\end{minipage}\hspace{0.018\textwidth}%
\begin{minipage}{0.3\textwidth}
\includegraphics[width=1.25\textwidth]{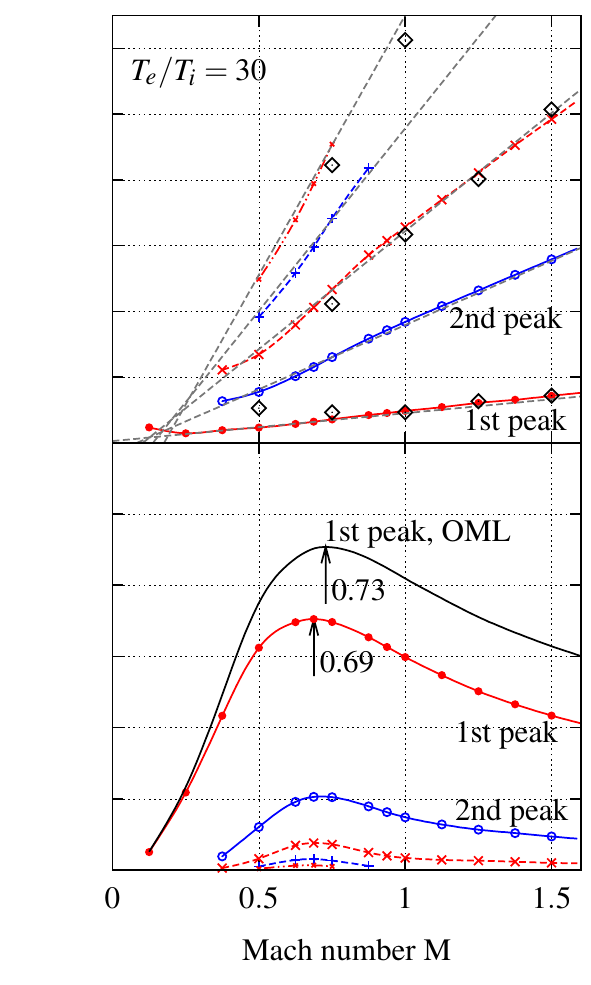}
\end{minipage}\hspace{0.018\textwidth}%
\begin{minipage}{0.3\textwidth}
\includegraphics[width=1.25\textwidth]{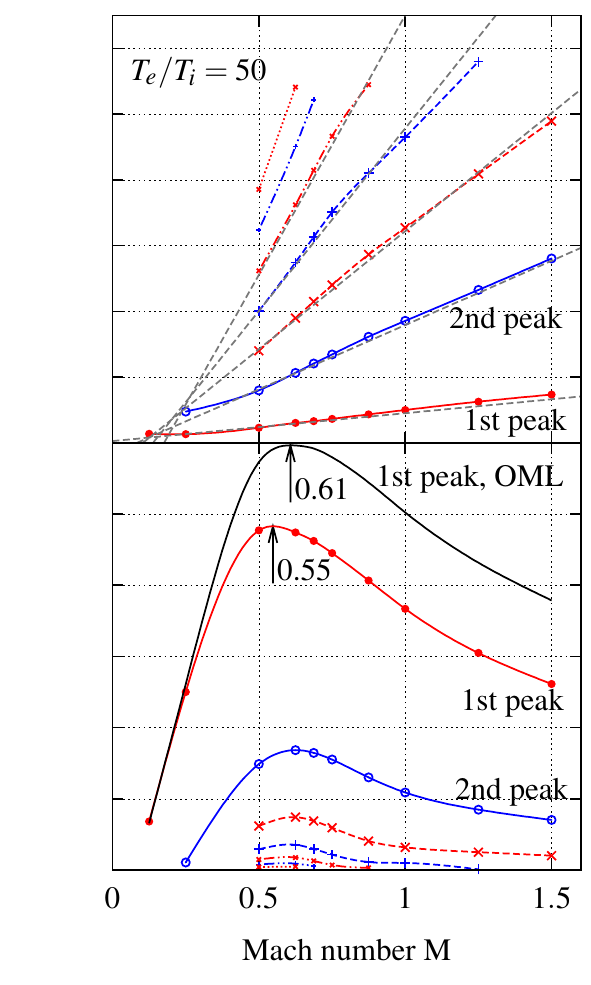}
\end{minipage} 
\caption{\label{pic:peakoverM} Peak positions (top) and peak amplitude (bottom) of the wake potential for $T_e/T_i=10,30,50$ (left to right panel) as a function of the Mach number $M$. Red (blue) curves correspond to  a positive (negative) space charge. The data points are interpolated by splines.
In the top (bottom) panel, the peaks in the wake are counted from the bottom (top) curve up (down).
The PIC results from Tab.~\ref{tab:PIC} for the wake maxima are denoted by diamond symbols.
The dashed black curves in the bottom left panel indicate the charge variation with $M$ for $T_r=10,30,50$ according  to OML theory (in arb. units.). 
The black solid curves (bottom) indicate the height of the first peak in the wake when the effect of charge variation is included.
}
\end{figure}

The potential profile can be studied in more detail in Fig.~\ref{pic:sideview}, where cuts through the potential surface are plotted for $M=0.375,0.6875,1.0$ and $T_{\rm r}=30$. For the considered grain charge of $Q_d= -10^4 e_0$ the potential height takes a value around 
$10^{-2}\cdot \kB T_e/e_0\approx25mV$, which is consistent with Refs.~\cite{hebner,hutch2011}.
Evidently, an oscillatory wake is not only formed when the speed of the ion flow exceeds the ion-acoustic velocity~\cite{vlad11,vladimirov}. Rather, a significant ion focus with a potential peak height of $21.7\,\mathrm{mV}$ is already found for $M=0.375$, i.e., well below ion sound speed. In fact, the highest amplitude of the wakefield is reached at $M=0.6875$, see also Fig.~\ref{pic:peakoverM}. 
In the upstream direction, the range of $\Phi_{\rm para}$ is already significantly enhanced for $M=0.375$, while in the cross-stream direction $\Phi_{\rm ortho}$  is still relatively close to the corresponding Yukawa potential for that value. When $M$ is increased, the (predominantly repulsive) potential profile in the upstream and cross-stream direction approach each other, as discussed in the context of Eq.~(\ref{lambda_s}).

In the following, we will discuss the functional characteristics of the wakefield downstream from the grain.
The potential peak height and its position as a function of $M$ are plotted in Fig.~\ref{pic:peakoverM}. 
The positions of the maxima and minima of the wake potential are found to shift linearly away from the grain, when $M$ is increased. Averaging over the three temperature ratios $T_e/T_i=10,30,50$ the best linear fit $a_i M + b_i$ for the $i$-th peak is obtained for the parameters $a_i=(0.848,3.91,7.19,11.1,15.8)\,\mathrm{mm}$ and $b_i=(0.0553,-0.327,-0.745,-1.54,-2.78)\,\mathrm{mm}$, see dashed gray lines in Fig.~\ref{pic:peakoverM} (top). Corresponding PIC results for the wake maxima (see Table~\ref{tab:PIC} in section~\ref{sec:PIC}) are marked with diamond symbols. Despite the finite, and relatively large grain size considered in the PIC simulations, there is a good agreement of both methods. On the basis of that result, a longitudinal wavelength of the wake equal to $2 \pi\lambda_{D_e} M$, discussed in Ref.~\cite{hutch2011}, holds in the relevant range of $M$ as a reasonable approximation.



\begin{figure}[t]
\begin{minipage}{0.475\textwidth}
\includegraphics[width=1.0\textwidth]{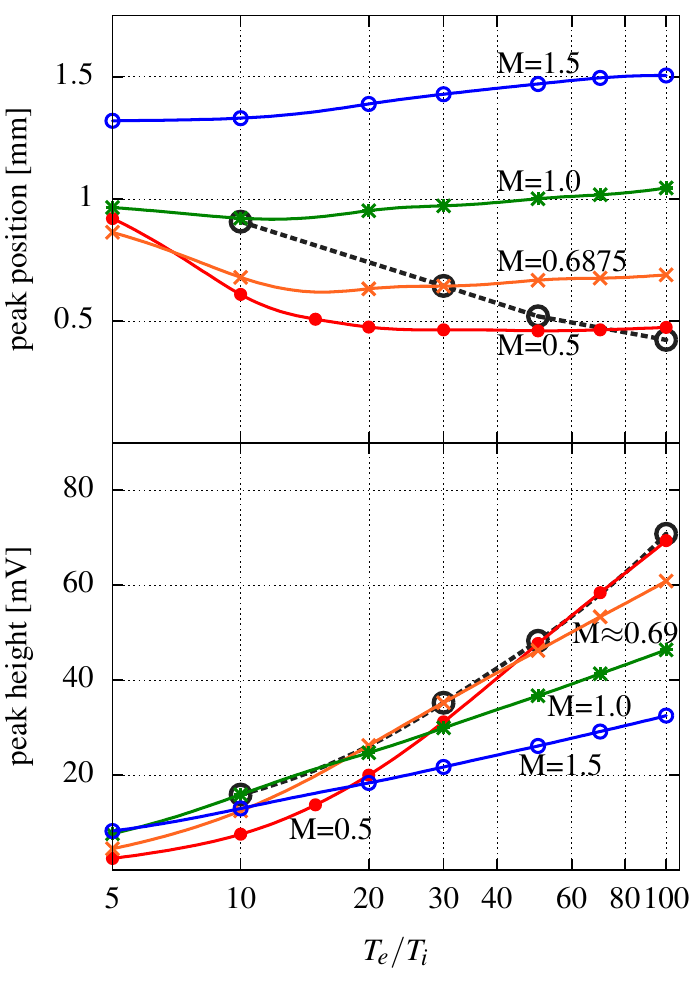}
\end{minipage}\hspace{0.04\textwidth}%
\begin{minipage}{0.475\textwidth}
\includegraphics[width=1.0\textwidth]{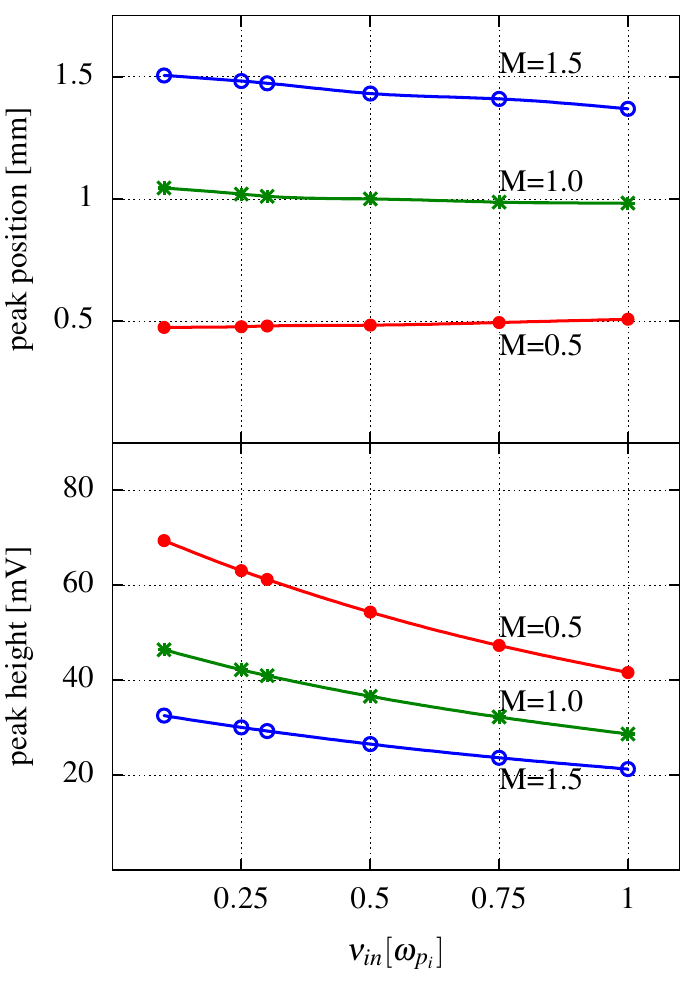}
\end{minipage}
\caption{\label{pic:peakoverTandnu} Positions (top) and heights (bottom) of the first peak of the wake potential for different $M$ values as a function of temperature ratio $T_e/T_i$ (left panel, $\nu_{in}=0.1$) and collision frequency $\nu_{in}$ (right panel, $T_e/T_i=100$). 
The data points are denoted by symbols and interpolated by splines. 
Open black circles mark the absolute maximum of the peak height as a function of $M$ (bottom left), cf. Fig.~\ref{pic:peakoverM}, and its position (top left) for different ratios of $T_r$.
}
\end{figure}

%

In Fig.~\ref{pic:peakoverM}, we also show the magnitude of the potential peak heights as a function of $M$.
As it is discussed in the context of Fig.~\ref{pic:sideview}, the wake amplitude reveals a clear maximum when $M$ is varied.  Interestingly, this maximum shifts to the subsonic regime when the ratio $T_e/T_i$ is increased (in Fig.~\ref{pic:peakoverM}, an arrow indicates the spline interpolated maximum). According to the OML theory \cite{fortov}, the dust charge $Q_{OML}^{T_r}$ increases monotonically with $M$ as shown for three temperature ratios in the bottom left panel in Fig.~\ref{pic:peakoverM} (see the dashed black curves). 
Taking the charge increase into account, the maximum value of the peak height shifts to slightly higher values of $M$ as displayed for the first potential peak (black solid line). 

The left panel of Fig.~\ref{pic:peakoverTandnu} shows the temperature dependence in more detail. 
For high ion temperatures, i.e. $T_r\leq5$, the largest peak amplitude is found in the supersonic regime, i.e. $M\geq1.5$. Increase of $T_r$ leads to a monotonic rise of the potential height (the effect of Landau damping is reduced). The greatest slope of the peak height with respect to $T_r$ is observed for $M=0.5$. So that in the limit of a large temperature ratio (a low ion temperature), the largest wake amplitude is found in the subsonic regime, that is $M=0.5$, whereas for the largest Mach number, $M=1.5$, the amplitude of the wake oscillation is considerably smaller. 
The maximum value of the first peak's height as function of $M$ and its position are denoted by open black circles for $T_r=10,30,50,100$.
The spline through these data points reveals that the strongest first peak in the wake (the primary potential maximum) over the full range of Mach numbers
shifts with increase of $T_r$ closer to the dust grain (top left panel). We note that a similar trend, as observed for the fixed grain charge (black dashed line), is found when one includes the effect of the charge increase according to OML theory (not shown). The right panel of Fig.~\ref{pic:peakoverTandnu} displays that an increase of ion-neutral scattering $\nu_{in}$ leads to an increased damping of the wake oscillation with the peak positions being changed only slightly.

\begin{figure}[t]
\hspace{-0.02\textwidth}
\includegraphics[width=1.0\textwidth]{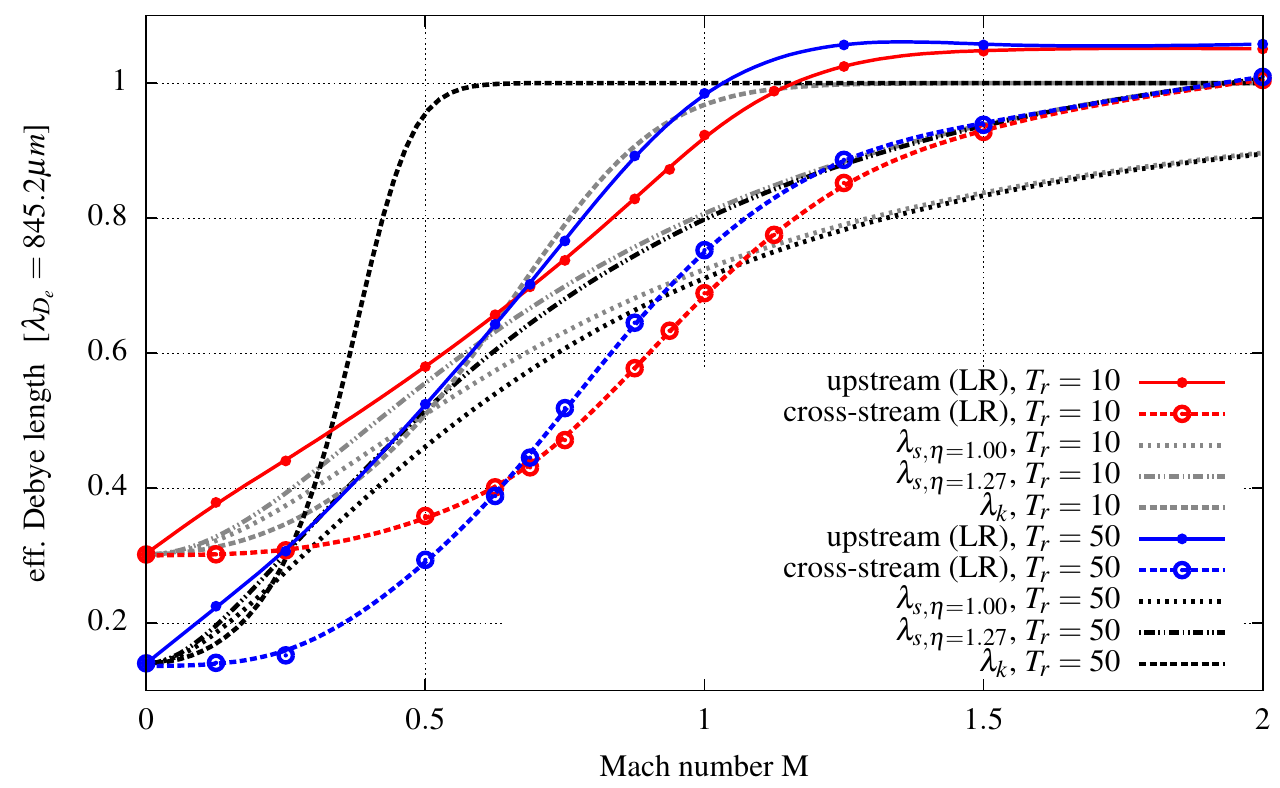}
\caption{\label{pic:debyeeffLR} Linear response (LR) results for the effective screening length in the upstream (solid) and cross-stream (dashed) direction relative to the grain as a function of the ion flow velocity~$M$. The LR results are given for two temperature ratios $T_r=10$ (red) and $T_r=50$ (blue). The light gray [black] lines indicate the characteristic screening length $\lambda_s$, Eq.~(\ref{lambda_s}), for $\eta=1.0, 1.27$, and $\lambda_k$, Eq.~(\ref{lambda_kraphak}), for $T_r=10$ [$T_r=50$], respectively. 
}
\end{figure}

So far, we mainly focused on the potential's wake. The form of the potential upstream and laterally from the grain can be captured by an effective screening length~\cite{lampe2000,hutch2006,jenko2005,konopka2000,khrapakPoP2005}. Fig.~\ref{pic:debyeeffLR} displays the Debye length as function of $M$ for different temperature ratios $T_r$.
For low ion drift, $u_i \ll v_{T_i}$, both electron and ions provide shielding, but ion screening typically dominates since $T_i<T_e$. For zero flow, $M=0$, the isotropic Yukawa potential, Eq.~(\ref{Yukawa}), applies with the total Debye length being $\lambda_D=0.30,0.18, 0.14$ in units of $\lambda_{D_e}$ for $T_r=10,30,50$, respectively. In this static case, the negatively charged dust grain Coulomb potential is predominantly screened by a positively charged ion cloud which isotropically forms around the grain. 
In contrast, in the supersonic limit $M \gg1$, (spherically symmetric) Debye screening is entirely contributed by the electrons.\footnote{See also the discussion of Eq.~(\ref{dk}) in section~\ref{theoapproach}.}
For finite $M$, the characteristic screening length is found to be different in the upstream and cross-stream direction, with the screening in the upstream direction being larger, see Fig.~\ref{pic:debyeeffLR}. 
It needs to be mentioned, that the characteristic screening lengths were obtained by fitting the analytic Yukawa potential to potential cuts through the dust grain such as shown in Fig.~\ref{pic:sideview}. However, these potential cuts do not exactly match the functional form of the Yukawa potential. For instance, for $M=0.375$ (left panel in Fig.~\ref{pic:sideview}) it is found that the orthogonal slice slightly overshoots $\Phi=0$, i.e. the potential possesses a weakly attractive branch sideways from the grain\footnote{The attractive part is found in a small parameter range only. In experiments and PIC simulations this weakly pronounced effect has not been observed so far~\cite{konopka2000}.}. 
Therefore, the screening lengths in Fig.~\ref{pic:debyeeffLR} cover the form of the potential close to the grain, but have an approximate character only. Also, a systematic error is observed in the limit $M \rightarrow \infty$, where the value of the effective screening length in upstream direction converges towards a value that is about five percent larger than the appropriate screening length that is the electron Debye length $\lambda_{D_e}$. 
In general, we find that the Yukawa potential matches the potential profile perpendicular to the flow essentially better than in the upstream direction which is in accordance with~\cite{konopka2000}.

Finally, we like to point out that the characteristic screening length can be roughly approximated by the interpolation formula 
\begin{align} \label{lambda_s}
\lambda_s= \lambda_{D_e} \left(\frac{k_B T_i +\eta m_i u_i^2}{k_B T_e + k_B T_i +  m_i  u_i^2}\right)^{1/2} = 
\left( \frac{1}{ f_s(\eta) \lambda_{D_i}^2} + \frac{ f_s(1)}{ f_s(\eta) \lambda_{D_e}^2}\right)^{-1/2}   \quad,
\end{align}
where $f_s(\eta)=1+\eta M^2 \frac{T_e}{Ti}$ and $\eta$ is a heuristic fitting parameter which we introduce here. Considering $\eta=1$, $\lambda_s$ reproduces both screening limits correctly: (i) $\lambda_s=\lambda_{D}$ for $M=0$, and (ii) $\lambda_s=\lambda_{D_e}$ for $M\rightarrow \infty$ (see the dotted lines in Fig.~\ref{pic:debyeeffLR}) \cite{hutch2006}. However, $\lambda_s$ is found to converge much slower towards $\lambda_{D_e}$ than the curves obtained by LR theory (Fig.~\ref{pic:debyeeffLR}) and PIC simulations (as will be shown in section~\ref{PICresults}, Fig.~\ref{fig:picdebye}). Therefore, it is suitable to introduce the fitting parameter $\eta=1.27$ which allows for an essentially better agreement with the present results in the relevant range $0\leq M \lesssim 2$.\footnote{At larger values of $M$ screening by the electrons plays the dominant role and the effective screening length is therefore close to $\lambda_{D_e}$.}
The concept of an effective screening length in a plasma with streaming ions was also discussed by S.A. Khrapak \textit{et al.}, see \cite{khrapakPoP2005} and on page 45 in Ref.~\cite{fortov2005}.  Here, the analytical form of the effective screening length is approximated by the formula
\begin{align} \label{lambda_kraphak}
\lambda_k = \left(f_{1(2)}(M_T)  \lambda_{D_i}^{-2} +  \lambda_{D_e}^{-2}    \right)^{-1/2} \quad,
\end{align}
where the two fitting functions  $f_{1}=\exp(-M_T^2/2)$ and $f_{2}=(1+M_T^2)^{-1}$  are defined in terms of the \textit{thermal} Mach number $M_T\equiv u_i/ v_{T_i}=M c_s/ v_{T_i}$.  Evidently, with the fitting function $f_2$, Eq.~(\ref{lambda_kraphak}) is identical to Eq.~(\ref{lambda_s}) with $\eta=1$ and is therefore not further considered in the following.
For $T_r=10$, Eq.~(\ref{lambda_kraphak}) yields in conjunction with $f_1$ a very reasonable approximation (especially for the screening in the upstream direction) without any heuristic parameter, see dashed gray line in Fig.~\ref{pic:debyeeffLR}. Furthermore, $\lambda_k$ shows (in contrast to $\lambda_s$) the correct analytic trend of 
a rapidly increasing slope 
with increasing temperature ratio $T_r$, which is evident in the intersection of curves at different temperatures.  However, $\lambda_k$  overestimates the decay of the ion contribution to the screening as shown for $T_r=50$ in Fig.~\ref{pic:debyeeffLR} and for $T_r=30$ in Fig.~\ref{fig:picdebye}.

Theoretical findings that the effective screening length converges toward the electron Debye length $\lambda_{D_e}$ as $M$ increases is supported by experimental results~\cite{arnas2001,konopka1997}. In those experiments the effective screening length for the grains suspended in the sheath was found by analyzing collisions between the dust grains. In both cases it was shown that supersonic ions are unable to screen the dust particles and therefore the electron Debye length is the appropriate screening length for dust grains in supersonic plasma flows.

\section{3D Particle-In-Cell Simulations} \label{sec:PIC}
The LR theory, employed in the previous section, neglects some important processes associated with the dust grain charging and wake formation. 
In particular, in the regime of strong plasma flows and high grain charges, non-linear effects can be significant. Particle kinetic simulations, such as the PIC method, allow for self-consistent studies of the grain charging, and account also for non-linear phenomena.

\subsection{Theoretical Approach} \label{pictheory}
Analytical plasma models that account also for non-linear effects are difficult to develop. Therefore, one often employs numerical kinetic particle simulations, in which plasma particle trajectories are followed in self-consistent force fields without many approximations, allowing for a self-consistent evolution of the system. As plasma particle trajectories are the characteristics of the Vlasov equation, the PIC simulations can be considered as an alternative way of solving the Vlasov equation for arbitrary particle distributions \cite{buchner2007}. 

Plasma simulations, in which forces are evaluated between all pairs of simulation particles, are typically very expensive in terms of computational time and memory. A simulation of $n$ particles, with such direct force calculations, has complexity $\mathcal{O}(n^2)$, as approximately $n^2$ equations need to be solved to find the forces on all the particles. $\mathcal{O}$ relates to the computer resources usage time. As the number of simulated particles needs to be large, especially in 3D simulations, such an algorithm would be very inefficient.

Introducing a grid {within the PIC scheme} can significantly reduce the complexity of the algorithm \cite{birds1991}. In the PIC method, the physical quantities of each simulation particle {are} weighted to neighboring grid points to build appropriate density fields on the grid. In the electrostatic approximation, the electric potential is found from the charge density on the grid points by solving the Poisson equation. Thus, the complexity of the PIC algorithm is $\mathcal{O}(n)+\mathcal{O}(n_g \log  (n_g))$, where $n_g$ denotes the number of grid points with $n_g \ll n$. $\mathcal{O}(n_g \log  (n_g))$ is the complexity of solving the field equations. 

Integration of particle trajectories puts limits on PIC simulations. With both electrons and ions simulated, the time resolution should be a fraction of the electron plasma period. The charging time is usually of the order of the ion plasma period, and thus it requires a large number of time steps. Simulations can be accelerated by assuming Boltzmann distributed electrons, which leads to a hybrid PIC-fluid code  \cite{Cartwright2000}. This approximation is however not always valid, as trapped electrons, or electron sources (e.g., due to photoemission) can give rise to local deviations from the Boltzmann distribution \cite{miloch2009njp}. Another way of speeding up the evolution of the system is by using a reduced ion mass $m_i$, as the ion plasma frequency increases with decreasing $m_i$. Ion mass values as low as  $m_i = 30m_e$ have been used in the literature \cite{spitk2008}.
In~\cite{miloch2007} it has been demonstrated that mass ratios $m_i/m_e > 100$ give reliable results for the dust charging. While the reduction of ion mass leads to some quantitative differences, the results are qualitatively correct, scalable and can be presented in normalized units. 
 Note that with the ion mass being reduced, the sound speed $c_s$ is increased and the floating potential is slightly reduced.
In turn, with a relatively large grain, the discreteness noise and charge fluctuations on the grain are diminished.

\subsection{Numerical implementation}
In the present study, we use the \textit{Dust in Plasma 3D} (DiP3D) particle-in-cell code, which has been developed to study the charging of dust grains and other objects in various plasma environments~\cite{miloch2010, miloch2010a}. The code simulates a collisionless plasma in the electrostatic approximation, with electrons and ions represented as individual plasma particles. To weight particle charges to the regular grid, and to project forces from the grid to the particles, first order linear weighting is used. The particle trajectories are advanced with the {\it leap frog} method characterized by a staggered time mesh \cite{birds1991}:
\begin{align} 
x_i(t+\Delta t)   &= x_i(t)+v_i(t+\Delta t/2)\Delta t \quad, \nonumber \\
v_i(t+\Delta t/2) &= v_i(t- \Delta t/2) + f_i(t) \Delta t /m_i \quad, 
\end{align}
where $i$ refers to an electron or ion, $f_i$ is the force projected on the $i$-th particle from the nearest grid points, and $\Delta t$ is the computational time step.

The plasma particles are initially uniformly distributed with Maxwellian velocity distributions in the simulation box of size $L^3=(24 \lambda_{{D_e}})^3$. The particles are injected into the simulation box at each time step according to prescribed particle fluxes. To simulate an open plasma system, the particles can leave freely through the boundaries of the simulation box. The code uses Dirichlet boundary conditions for the potential. The potential at the boundaries is set according to the plasma potential $\Phi_{pl}= 0$ V. 

One or several spherical dust grains are placed inside the simulation domain far away from the boundaries.  We assume a conducting grain, and each plasma particle that hits the surface is removed and contributes to the current to the grain. The charge of the removed particle is distributed on the grain surface to cancel internal electric fields. The grain is initially uncharged and becomes charged self-consistently during the simulation.  

The DiP3D code is run on a computer cluster, with typically $n=10^7$ or more simulation particles distributed on several nodes, with a typical grid size of $n_g\geq128^3$. The Message-Passing-Interface (MPI) is used for parallelization. The code can be stopped and then restarted with modified dust grain configurations, which can reduce the computation time in case of systematic parametric studies. The code also allows for a movement of the simulated object, and for the inclusion of various charging processes such as photoemission. Recently, the code has been upgraded to include external magnetic fields and ion-neutral and electron-neutral collisions \cite{miloch2010b}. With all the features, the code can be placed among other cutting-edge particle plasma codes for object-plasma interactions \cite{hutch2006, matyash2010, engw2006}. In the present work, we make only use of the basic features of the code.

It should be noted that the dimensionality of the code is an important aspect of any simulation. For the ultimate goal of direct application of the numerical results to support laboratory studies, 3D simulations should be used. Many important effects, such as ion focusing, are still present when the dimensionality is reduced. 
However, in a 2D system, the charge of the plasma particle corresponds to the charge of an infinite rod, and a simulated circular grain represents an infinite cylinder. 
The plasma frequency $\omega_p$, Debye length $\lambda_D$, and charge density do not change with the dimensionality of the system. Reducing the dimensionality can give faster algorithms, but while the results of such simulations are qualitatively correct, they do not need to represent a 3D system quantitatively. Thus, results from codes with reduced dimensionality should be compared with corresponding theoretical models. In the present paper, we present results from 3D simulations.


\subsection{Results} \label{PICresults}

\begin{figure}[t]
\includegraphics[width=\textwidth]{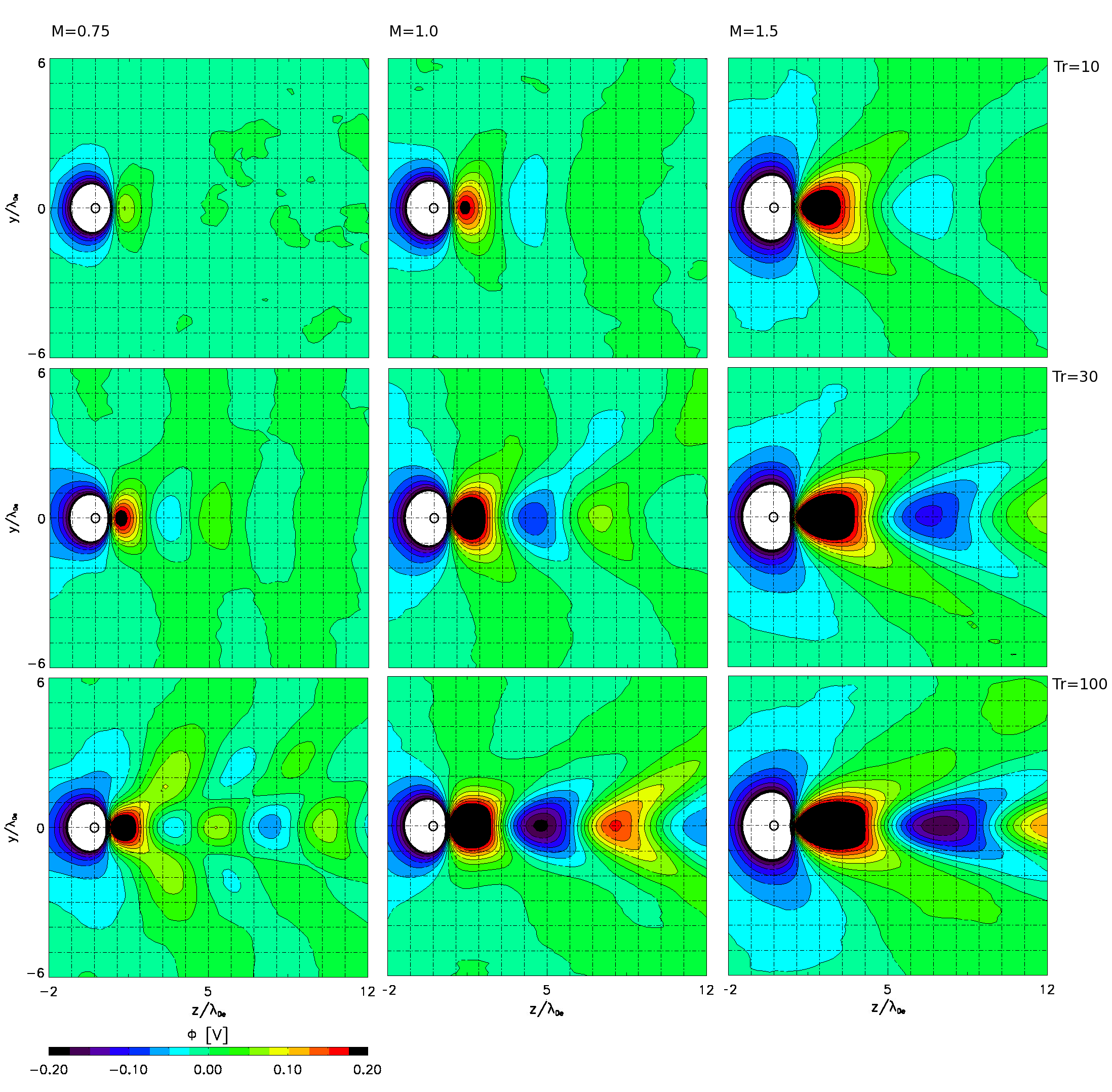}\hspace{2pc}%
\begin{minipage}[b]{\textwidth}\caption{\label{fig:PIC9} First principle results from collisionless 3D PIC simulations. The contour plots show the wake-potential $\Phi$ in the $y$-$z$ plane at $x=0$  for varied ion streaming velocities $M=0.75, 1.0, 1.5$ (from left to right) along $z$ axis, and electron-ion temperature ratios $T_r=10, 30, 100$ (from top to bottom). The finite-sized dust grain is placed at the origin.  Blue (red) color codes denote areas of negative (positive) space charge; potential values larger than $\Phi=0.2\,\mathrm{V}$ (lower than $\Phi=-0.2\,\mathrm{V}$) are clipped and shown black (white). The size of the dust grain is marked by the black, open circle at the origin of the coordinate system.}
\end{minipage}
\end{figure}

With the DiP3D code, we performed simulations of a single grain in a collisionless plasma. The grain radius is $a=0.185\lambda_{{D_e}}=0.1563\,\mathrm{mm}$.
The finite size of the grain leads to enhanced potential variations in the wake as compared to the previous LR results, because the self-consistent grain charge is on the order of $Q_d \approx -10^5e_0$. 
All other plasma parameters are as described in section~\ref{plasmaparameters}.  The grain charging and wake formation have been investigated for different 
electron-to-ion temperature ratios, $T_r \in \left\{10,30,100\right\}$, and 
Mach numbers, $M  \in \left\{0.75,1,1.5\right\}$, as shown in the contour plots of the electric potential in  Fig.~\ref{fig:PIC9}.
In all cases, we observe a positive maximum in the potential distribution in the wake at a distance of approximately one to two $\lambda_{{D_e}}$ downstream from the dust grain. This potential maximum is related to the ion focus in the wake of the negatively charged grain. The maximum in the potential is located further downstream from the grain than the enhancement in the ion density in the focus region, which is consistent with the Poisson equation $\nabla^2 \Phi=-\rho/\epsilon_0$, where $\rho$ is the charge density. Ion focusing is due to the deflection of ion trajectories by electric fields in the vicinity of a negatively charged grain~\cite{miloch2008, schw1996}.

\begin{figure}[t]\hspace{-0.01\textwidth}%
\begin{minipage}{0.3\textwidth}
\includegraphics[width=1.25\textwidth]{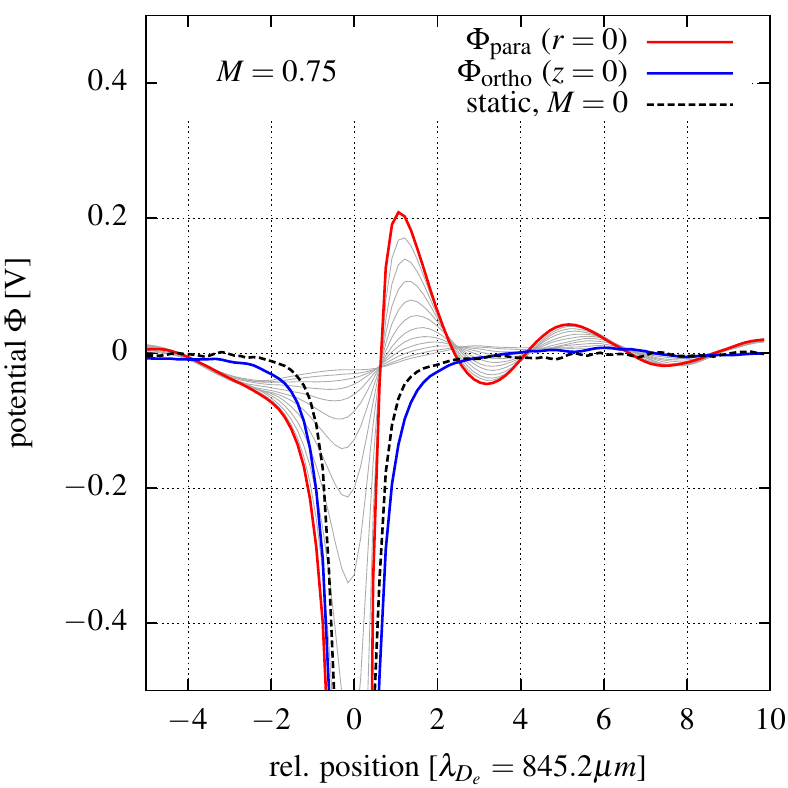}
\end{minipage}\hspace{0.018\textwidth}%
\begin{minipage}{0.3\textwidth}
\includegraphics[width=1.25\textwidth]{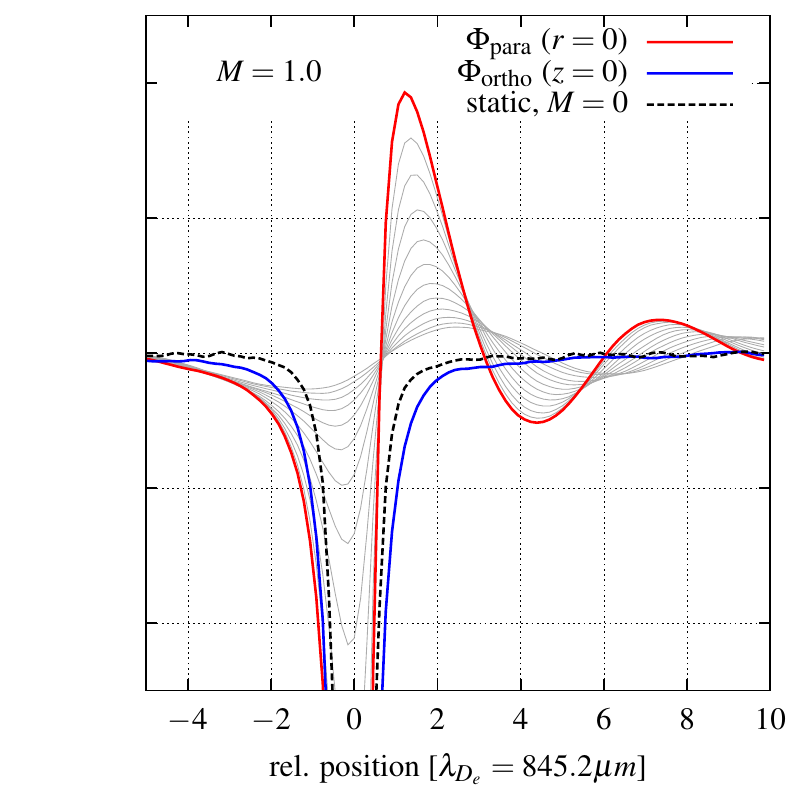}
\end{minipage}\hspace{0.018\textwidth}%
\begin{minipage}{0.3\textwidth}
\includegraphics[width=1.25\textwidth]{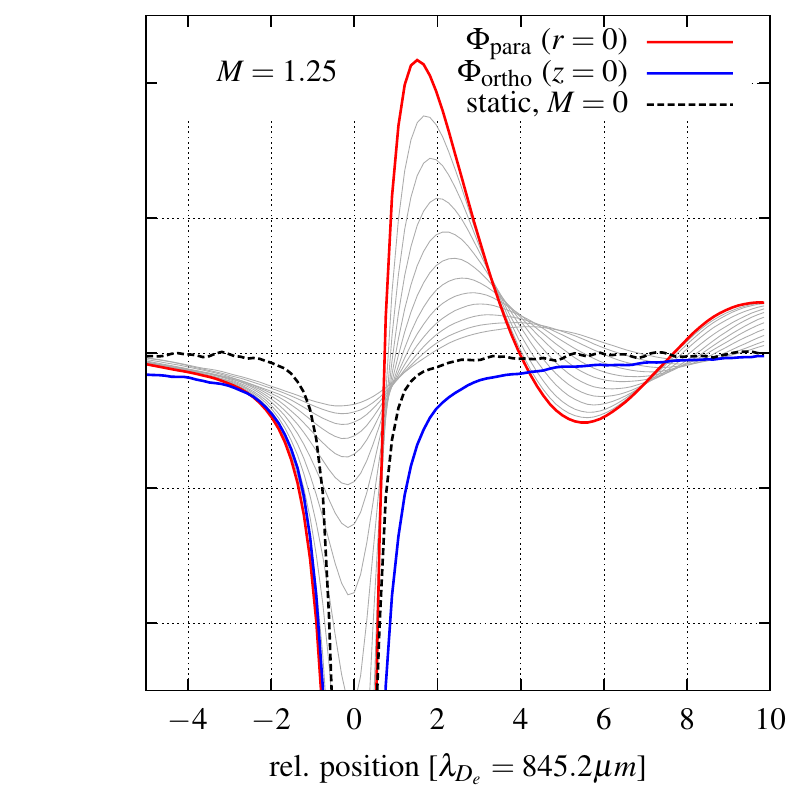}
\end{minipage}
\caption{\label{fig:cutPIC3} 3D PIC results for  $T_e/T_i=30$: potential along the flow direction $z$ through the centre of the grain $x=y=0$ (red line) and in the cross-stream direction at $z=0$ (blue line) for $M=$0.75, 1.0, and 1.25.  Light gray curves represent slices through the potential surface for $0 \leq y \leq 2\lambda_{D_e}$ (along the direction of the flow).
The dashed black line shows the potential drop for the static case, $M=0$, for which $\Phi_{fl}=-2.02$V. 
As the floating potential of the grain changes with the flow speed, the results for the static case have been rescaled, such that the potential on the grain matches the floating potential on the grain in a flowing plasma. 
}
\end{figure}

\begin{figure}[t] 
\begin{minipage}{0.532\textwidth}
\includegraphics[width=1.0\textwidth]{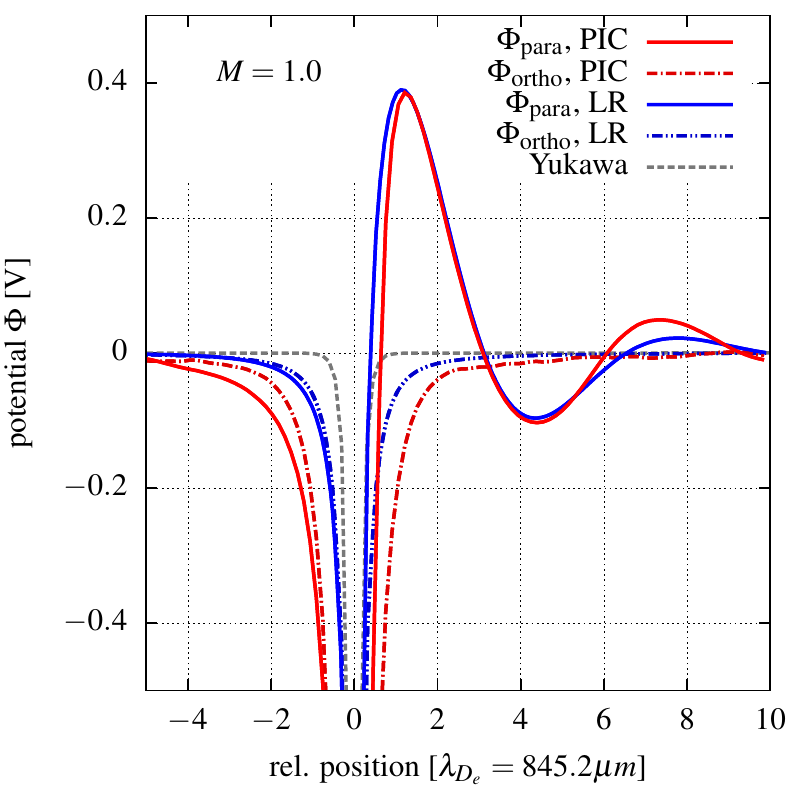}
\end{minipage}\hspace{0.025\textwidth}%
\begin{minipage}{0.4467\textwidth}
\includegraphics[width=1.0\textwidth]{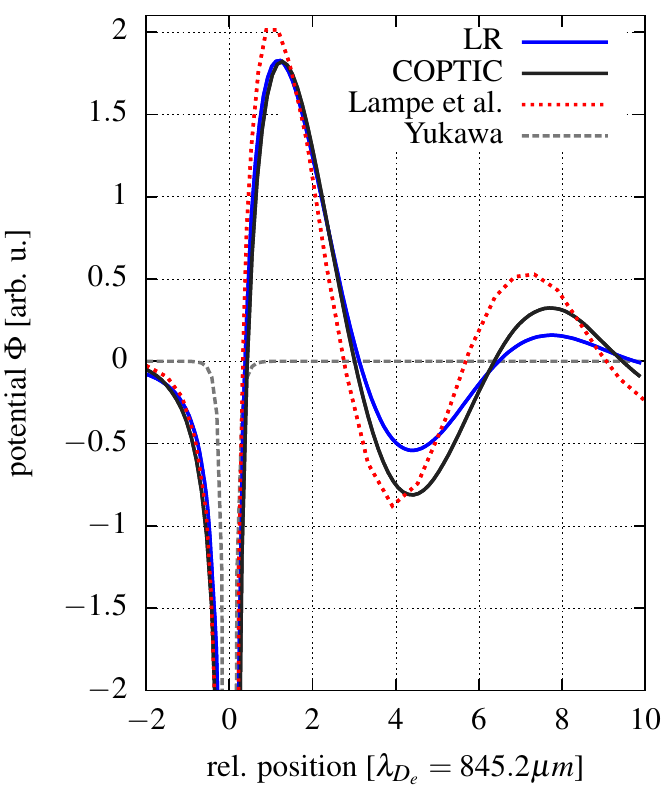}
\end{minipage}
\caption{\label{fig:PICvsLR} 
\underline{Left:} Potential cuts through the grain along the $z$-direction (solid lines) and in cross-stream direction (dash-dotted lines) for $M=1.0$ and $T_e/T_i=30$ as obtained from collisionless 3D PIC (red) and LR calculations with $\nu_{in}/\omega_{p_i}=0.1$ (blue). The grain charge of the LR result and the corresponding Yukawa potential are adjusted to match the potential of the PIC simulation, which includes the grain charging process self-consistently. Note that in contrast to PIC, the LR results do not take into account the finite particle radius $a=0.185\lambda_{D_e}$, which explains the offset of the curves. 
\underline{Right:} Potential cuts along the $z$-direction for $M=1.0$ and $T_e/T_i=50$. Shown are our LR calculations with finite damping included $\nu_{in}/\omega_{p_i}=0.1$ (blue solid line), COPTIC data by Hutchinson for a point-like grain in a collisionless plasma  (black solid line) \cite{hutch2011}, and linear response results for the collisionless case by Lampe \textit{et al.} (red dotted line)~\cite{lampe2000}.
}
\end{figure}

%

\begin{figure}[t]
\includegraphics[width=1.0\textwidth]{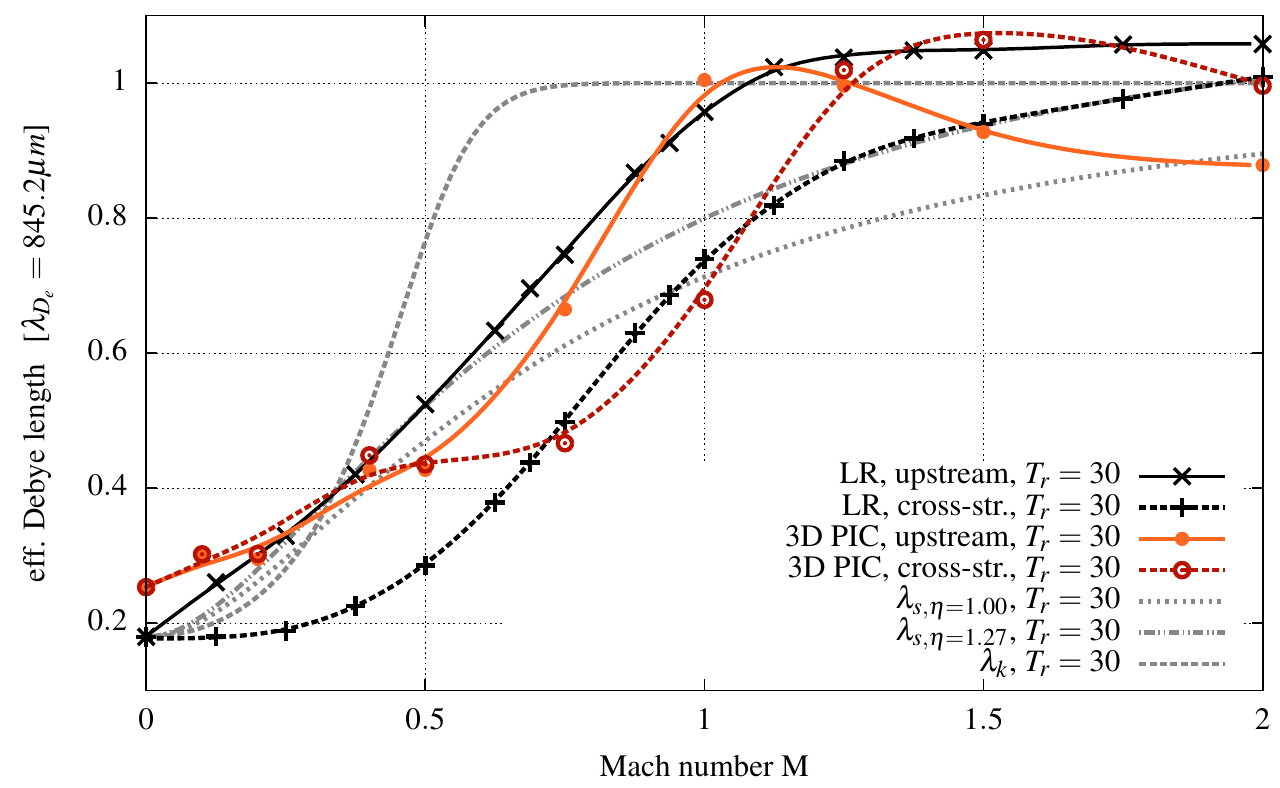}\hspace{2pc}%
\caption{\label{fig:picdebye} Effective screening length of the dust grain potential for $T_r=30$ as a function of the ion flow velocity $M$.
PIC [LR] results for the upstream (solid curve) and cross-stream (dashed) direction are indicated by circles [crosses] on orange/red [black] lines. These lines are result of an approximating spline interpolation and therefore do not exactly pass through the data points. The estimated error of the PIC data points is $\delta\lesssim0.05\lambda_{D_e}$. 
The light gray lines indicate the screening length $\lambda_s$, Eq.~(\ref{lambda_s}), for $\eta=1.0, 1.27$, and $\lambda_k$, Eq.~(\ref{lambda_kraphak}), respectively. See also Fig.~\ref{pic:debyeeffLR}.
}
\end{figure}

Similarly to the LR results in section~\ref{DSM}, in the PIC simulations the potential enhancements in the wake become more pronounced for larger $M$ and larger electron-to-ion temperature ratios $T_r$, i.e., colder ions. The first local potential maximum is followed by other potential extrema further downstream. 
For $T_r=100$ and $M=0.75$, the pattern in the wake shows a diverse structure with several minima and maxima being located also away from the flow symmetry axis. Note that this could also be an artefact due to the Poisson solver on the finite grid. In particular for very low $M$ and $T_r$, 
the possibility of instabilities can become an important issue for the convergence of the code \cite{Brackbill_1988}.
For larger $M$ and $T_r$, the extrema in the potential distribution are along the direction of the flow. With an increase of the ion temperature, the Landau damping becomes important, and thus ion focusing and wake patterns are less pronounced. For  $T_r=10$ and $M=0.75$ only a weak, single maximum is observed. The considerable structural consequences of even a weak ion focus were subject of recent investigations~\cite{ludwig2011}.

In Fig.~\ref{fig:cutPIC3}, we present potential cuts along the flow direction $z$ at $y=0$ and $x \in [0, 2\lambda_{{De}}]$ for $T_r=30$.
In the figure, we also plot the potential cut in the cross-stream direction, as well as the potential cut for the grain in stationary plasma. 
In contrast to the LR results, the potential maximum monotonically increases with the flow speed, and so does its extension in the perpendicular direction	whilst the Mach cone forms. However, there is an important difference between the LR and PIC results. In the LR analysis, the charge on the grain is constant, while in the PIC simulations it changes with the flow. The floating potential on a dust grain in a stationary plasma is $\Phi_\mathrm{fl}=-2.02\,\mathrm{V}$ and it becomes larger with increasing $M$ (see discussion below). While the charge and potential on the grain increase with the flow, the amplitude of peaks in the wake potential is being enhanced.

In order to compare the PIC results with LR calculations, in the left panel of Fig.~\ref{fig:PICvsLR}, we present potential cuts through the grain along the $z$-direction and in the cross-stream direction for $M=1.0$ and $T_e/T_i=30$ as obtained from collisionless 3D PIC simulations and from LR calculations with collisional damping included ($\nu_{in}/\omega_{p_i}=0.1$). The Yukawa potential, Eq. (\ref{Yukawa}), for a grain in stationary plasma is also shown. The grain charge of the LR result is adjusted to match the potential of the first peak in the PIC simulation (the same charge is used for the Yukawa potential). It can be inferred from the plot that the agreement between the PIC and LR results is good for the first two extrema. 
It is important to note that in contrast to the PIC simulations, the LR results do not take into account the finite particle radius $a=0.185\lambda_{{D_e}}$. For this reason, the LR curves are slightly shifted towards zero. When the finite size of the grain is taken into account, the agreement between the LR and PIC results for the potential profiles is improved.

As discussed in the context of Fig.~\ref{pic:peakoverM}, we find an overall good agreement in the longitudinal wavelength of the PIC and LR results.\footnote{In accordance with~\cite{schweig2005}, the linear approach tends to overestimate the asymmetry of the potential distribution around the dust particle.}
A more inconsistent discrepancy in the wavelengths was recently reported by Hutchinson~\cite{hutch2011}. 
Therefore, in the right panel of Fig.~\ref{fig:PICvsLR} a comparison for $T_e/T_i=50$ and $M=1.0$ is shown between our LR results, COPTIC results for a point-like particle (Fig. 11 in \cite{hutch2011}), and the linear calculation by Lampe \textit{et al.} (Fig. 3 in \cite{lampe2000}). The main difference between Lampe \textit{et al.}'s and our LR results is due to the fact that we include finite collisional damping (which in our case is necessary to avoid pseudoperiodicity effects for the given temperature ratio $T_e/T_i=50$, as stated in section \ref{numimplr}). The wavelength obtained from COPTIC is in full accordance with our LR calculation.\footnote{Collisional damping has on a minor effect on the peak positions, see right panel in Fig.~\ref{pic:peakoverTandnu}.}
The stronger decay of the wake oscillations can be attributed to finite collisional damping which is included in the LR calculations ($\nu_{in}/\omega_{p_i}=0.1$).
 However, in the left panel of Fig.~\ref{pic:peakoverM}, we find a much better agreement in the wake amplitude at the first subsidiary valley.
The stronger damping in our PIC simulation (left panel) must be due to the large grain size (and charge) giving rise to much stronger non-linear effects than the point-like grain in the COPTIC simulation. As discussed in \cite{hutch2011}, a major effect of non-linearity is the local increase of the effective ion temperature in conjunction with an enhanced Landau damping of the wake amplitude.
The presence of (strong) non-linearities in the present PIC (DiP3D) simulations becomes apparent by the similarity of the shape of the primary potential maximum for $M=1$ and the temperature ratios $T_r=30$ and especially $T_r=100$ (Fig. \ref{fig:PIC9}) and the non-linear wake structure presented in Fig.5(a) in Ref. \cite{hutch2011}. 
Therefore, we assume that the good agreement between our collisionless PIC simulation and the LR calculation with collisions included---not only in the wavelength of the wake oscillations but also in the decay of the wake amplitude---is due to a nonlinearity-induced damping mechanism that reduces the amplitude of the wake potential.


In the downstream direction, the primary repulsive branch of the potential is shifted closer towards the grain at finite (subsonic) flows, which is similar to the LR results (cf. Figs. \ref{pic:sideview} and \ref{fig:cutPIC3}).
As ions  contribute more to the grain screening in stationary than in flowing plasma, the grain screening is strongest in the static (Yukawa) case and weakens for finite $M$, see again Fig.~\ref{fig:cutPIC3}.
The quantitative comparison of the PIC and LR results has been done by calculating the effective screening length from the potential profiles for different $M$ for $T_r=30$ in the cross-stream and upstream directions, see Fig.~\ref{fig:picdebye}.
The general trend for the PIC and LR results is that the effective screening length $\lambda$ increases with the flow velocity $M$. The effective screening for the stationary plasma in the PIC simulations is $\lambda=0.23\lambda_{D_e}$. This value is larger than the the results from the LR calculations, as well as larger than the total Debye length, which is $\lambda_{D}=0.18\lambda_{D_e}$. The difference is due to a finite grid resolution in the simulation, where the grid spacing is $\Delta x \approx \lambda_D$. For the flow velocities $M < 0.5$, the cross-stream and upstream shielding lengths are lower than $0.4\lambda_{D_e}$. The values are roughly $0.1\lambda_{D_e}$ larger (smaller) than the cross-stream (upstream) results from LR calculations. For $M>0.5$, the effective screening length from the PIC simulations strongly increases with $M$, with steeper increase for the upstream direction. This is the same behavior as for the LR results. At $M \leq 1$, for both LR and PIC results, the screening length $\lambda$ is larger on the upstream than on the downstream side. The increase of the effective screening length with the flow is in agreement with previous studies \cite{lampe2000, jenko2005}, however, the details of the trend for cross-stream and upstream directions are different. In Ref.~\cite{lampe2000} the cross-stream shielding was systematically larger than the upstream one. Based on our results from simulations with two different numerical methods, the upstream shielding length is larger than the cross-stream one for $M \leq 1$. For higher $M$, the trend continues for the LR calculations, while for PIC simulations, the shielding becomes larger on the cross-stream than on the upstream side. The effective screening length converges towards the electron Debye length for supersonic flows, as it was discussed in the previous section for the LR results.

Finally, in Table \ref{tab:PIC}, we summarize the floating potential on the grain as well as the positions and heights of the maxima  for different $M$ values. Since the ion flux to the grain surface changes with the ion flow speed, the floating potential $\Phi_{fl}$ on the grain becomes more negative with increasing $M$. Changing $T_r$ has little influence on the floating potential, as $\Phi_\mathrm{fl}$ is mainly controlled by the electron temperature $T_e$, which is kept constant in our simulations. The heights of the potential maxima increase with the flow speed, and so does the distance between the peak and center of the grain. 

\begin{table}[!t]
\caption{Floating potential $\Phi_{fl}$ on the grain and the position $p_i$ and height $h_i$ of the $i$-th maximum in the wake. For some lower velocities only the first maxima were observed. Crosses ($\times$) correspond to possible maxima located outside the simulation domain.  The floating potential on a grain in a stationary plasma is  $\Phi_\mathrm{fl}=-2.02\,\mathrm{V}$ ($T_r=30$). Note that these results apply to the reduced ion mass. (In particular, the floating potential is therefore reduced, as discussed in section \ref{pictheory}).}
\label{tab:PIC}
\begin{eqnarray*}
\\T_e/T_i = 10\\
\begin{array}{rrrrrr}
\hline
M & \Phi_{fl} & p_1  & h_1 & p_2 & h_2 \\
  & [V] & [\lambda_{De}] & [V] & [\lambda_{D_e}] & [V] \\
  \hline
0.75 & -2.8 & 1.25 & 0.08 & - & - \\
1.00 & -3.1 & 1.25 & 0.195 & 7.5 & 0.02 \\
1.25 & -3.4 & 1.5	& 0.265	&8.5	& 0.01\\
1.50 & -3.6 & 1.75 &	0.3	& 11& 	0.25 \\
\hline
\hline
\end{array}
\newline \\
\\T_e/T_i = 30\\
\begin{array}{rrrrrrrr}
\hline
M & \Phi_{fl}  & p_1  & h_1 & p_2 & h_2 &p_3 & h_3 \\
  & [V] & [\lambda_{D_e}] & [V] & [\lambda_{D_e}] & [V] &[\lambda_{D_e}] & [V] \\
  \hline
0.50	& -2.4 & 1.25	& 0.06	& - & - &- & -\\			
0.75 & -2.8 & 	1.1 &	 0.22	 & 5	& 0.42	&10	&0.03 \\
1.00	&-3.1 &1.1&	0.35&	7.5	&0.6&	14.5&	0.04\\
1.25	& -3.3&1.5&	0.45&	9.5&	0.75&  \times&\times \\		
1.50	&-3.5 & 1.7&	0.45&	12	&0.65& \times&\times\\	
\hline
\hline
\end{array}	
\\T_e/T_i = 100\\
\begin{array}{rrrrrr}
\hline
M & \Phi_{fl}  & p_1  & h_1 & p_2 & h_2 \\
 & [V] & [\lambda_{D_e}] & [V] & [\lambda_{D_e}] & [V] \\
  \hline
0.50	&-2.3 & 1	&0.2&	3.5&	0.1 \\
0.75	& -2.8 &1.1&	0.33&	5.5	&0.07\\
1.00 & -3.2	&1.1&	0.55&	8	&0.165\\				
1.50& -3.8 &	1.7& 0.55	&12.2&	0.14 \\
\hline
\hline
\end{array}
 \end{eqnarray*}
\end{table}

\section{Method-Spanning Comparison} \label{sec:discussion}
We now summarize the correspondence of the two methods and discuss their limitations.

\subsection{Comparison of the potentials} 
The dynamical screening potentials obtained from the LR approach are topologically similar to the corresponding results from the PIC simulations, see again Figs.~\ref{pic:matrix} and \ref{fig:PIC9}. The PIC results are more noisy, as they account for the dynamics of plasma particles and fluctuations in the system. Since the PIC simulations consider a finite-sized grain that is self-consistently charged, the potential on the grain changes with the flow speed, see Table~\ref{tab:PIC}. The comparison of the peak positions of the positive ion wake charges, Fig.~\ref{pic:peakoverM}, shows a good agreement. Note that the exact position of weak peaks with a low signal to noise ratio (high peak numbers at small $T_r$ and $M$ values) 
can just barely be resolved, cf. Fig.~\ref{pic:matrix}.

The amplitude of the peak in the wake calculated with LR achieves a maximum at a certain ion flow velocity (Fig.~\ref{pic:peakoverM}).
This maximum is at lower flow velocities for larger temperature ratios. We do not see a maximum in the peak amplitude for the PIC results summarized in Table~\ref{tab:PIC}. However, in section~\ref{DSM}, we considered the charge on the grain to be constant, while in the PIC simulations the potential on the grain changes with the flow. When scaling linearly all PIC results to a fixed potential, we obtain a maximum in the height of the peak as a function of flow velocity. This maximum is at lower velocities for higher temperature ratios, similarly to the LR results. Moreover, adjusting the correct grain charge, the potential pattern in the wake from the LR simulations agrees well with the PIC results, see again Fig.~\ref{fig:PICvsLR}.

Both, LR theory and PIC simulations confirm that the effective screening of the grain changes with the flow. While the effective screening length calculations with the PIC method can be considered as less accurate due to potential representation on a finite grid, the trend for the effective screening length agrees with the LR results. For subsonic ion flow velocities, the screening length on the upstream side is usually larger than on the cross-stream side. 

\begin{figure}[t]\hspace{-0.01\textwidth}%
\begin{minipage}{0.33\textwidth}
\includegraphics[width=1.0\textwidth]{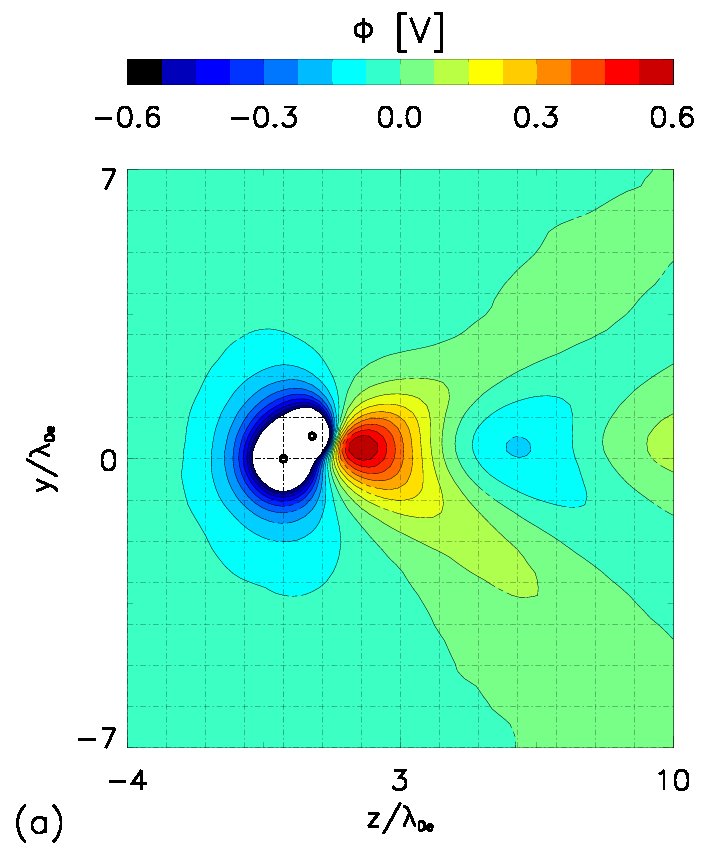}
\end{minipage}\hspace{0.018\textwidth}%
\begin{minipage}{0.33\textwidth}
\includegraphics[width=1.0\textwidth]{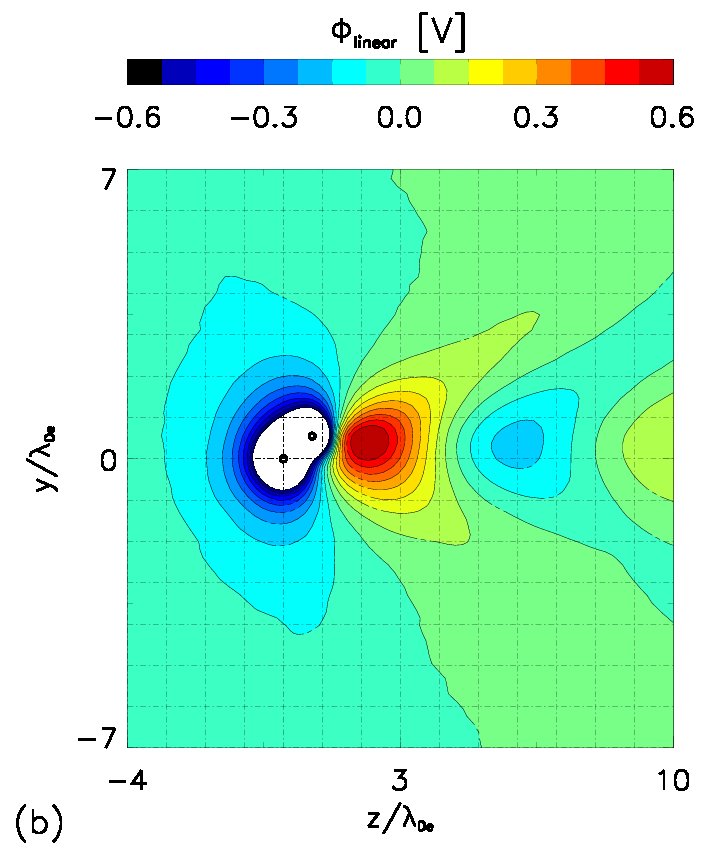}
\end{minipage}\hspace{0.018\textwidth}%
\begin{minipage}{0.33\textwidth}
\includegraphics[width=1.0\textwidth]{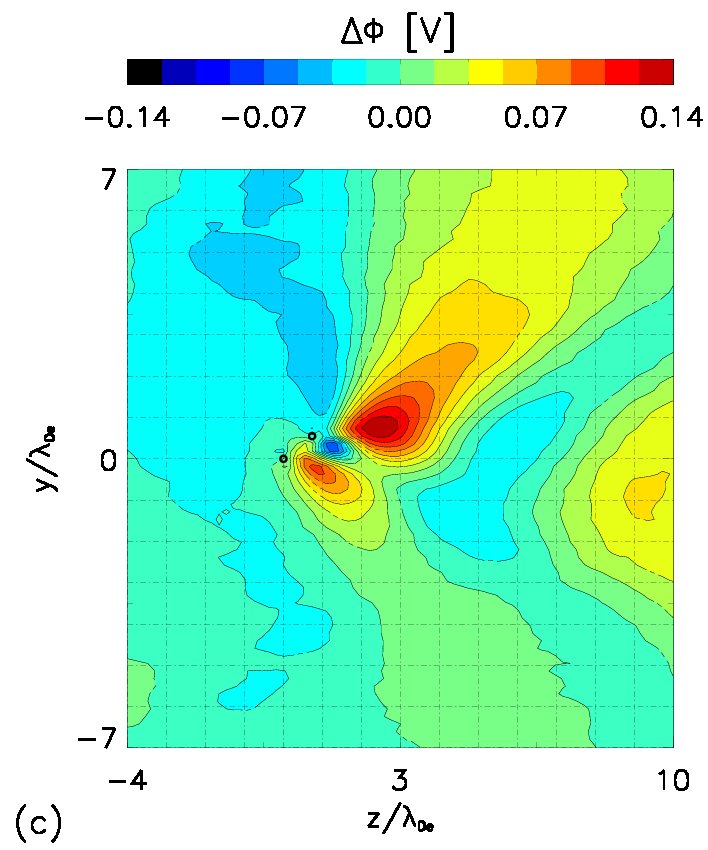}
\end{minipage} 
\begin{minipage}[b]{\textwidth}\caption{\label{fig:diff} (a) Potential around two grains separated by approximately $\lambda_{D_e}$, for $T_e/T_i=30$ and $M=1.25$ obtained by PIC simulations. (b) The potential constructed by a linear combination of the single grain wake potentials. (c) The potential difference between the constructed potential and the results from the simulation $\Delta \Phi= \Phi_{\mathrm{linear}}-\Phi$. Note the different values on the color bars.
}
\end{minipage}
\end{figure}

\begin{figure}[t]\hspace{-0.01\textwidth}%
\begin{minipage}{0.33\textwidth}
\includegraphics[width=1.0\textwidth]{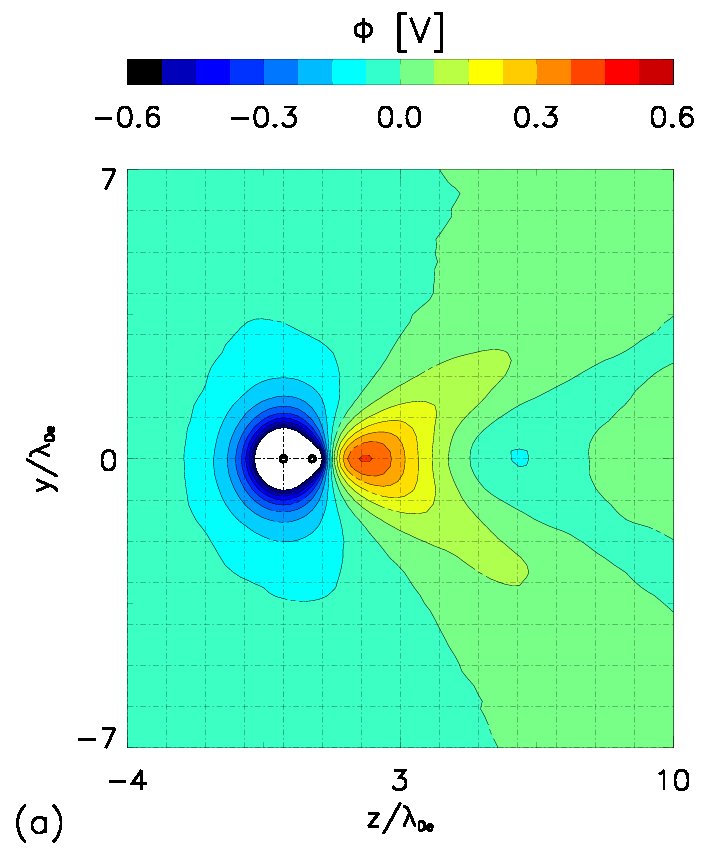}
\end{minipage}\hspace{0.018\textwidth}%
\begin{minipage}{0.33\textwidth}
\includegraphics[width=1.0\textwidth]{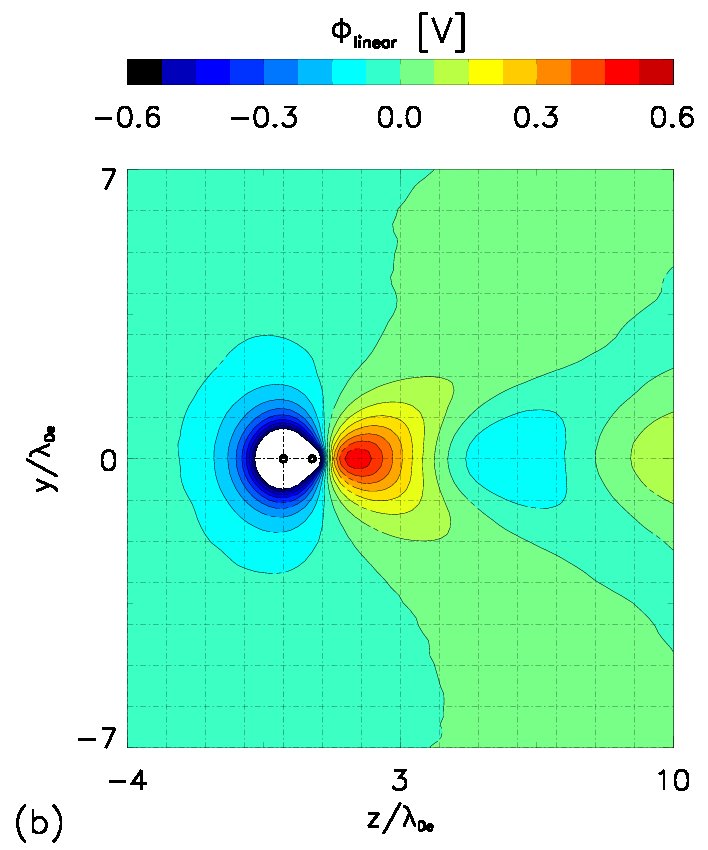}
\end{minipage}\hspace{0.018\textwidth}%
\begin{minipage}{0.33\textwidth}
\includegraphics[width=1.0\textwidth]{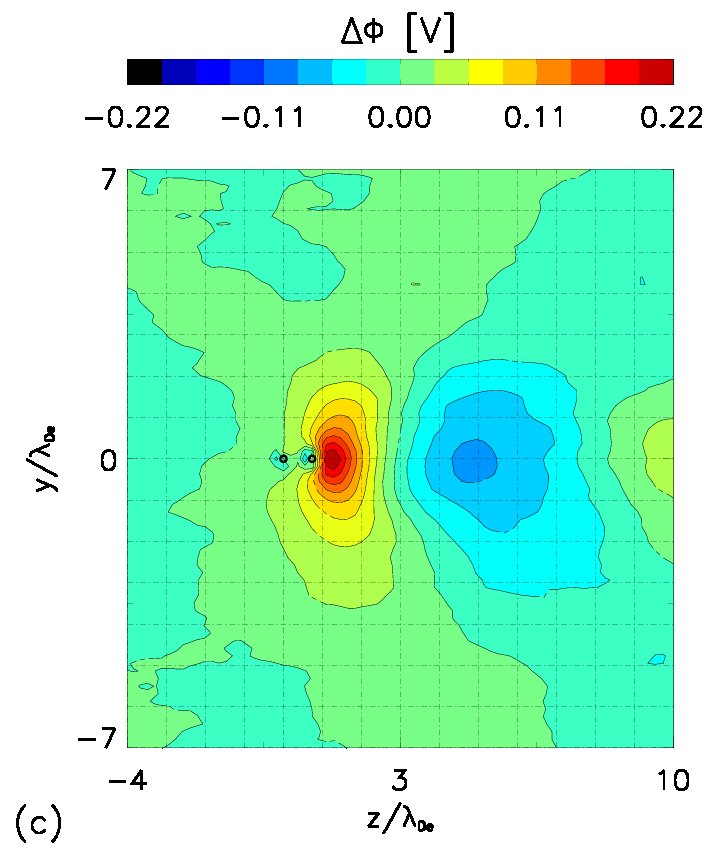}
\end{minipage} 
\begin{minipage}[b]{\textwidth}\caption{\label{fig:diff2} Same as Fig.~\ref{fig:diff}, but for two flow-aligned grains ($M=1.25$).
}
\end{minipage}
\end{figure}

\begin{figure}[h]\hspace{-0.01\textwidth}%
\begin{minipage}{0.33\textwidth}
\includegraphics[width=1.0\textwidth]{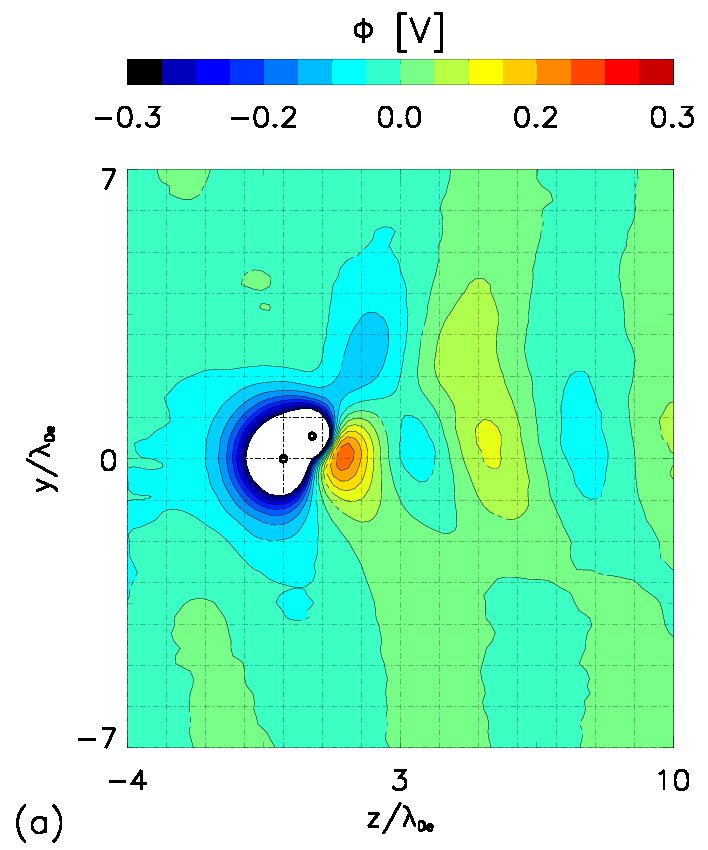}
\end{minipage}\hspace{0.018\textwidth}%
\begin{minipage}{0.33\textwidth}
\includegraphics[width=1.0\textwidth]{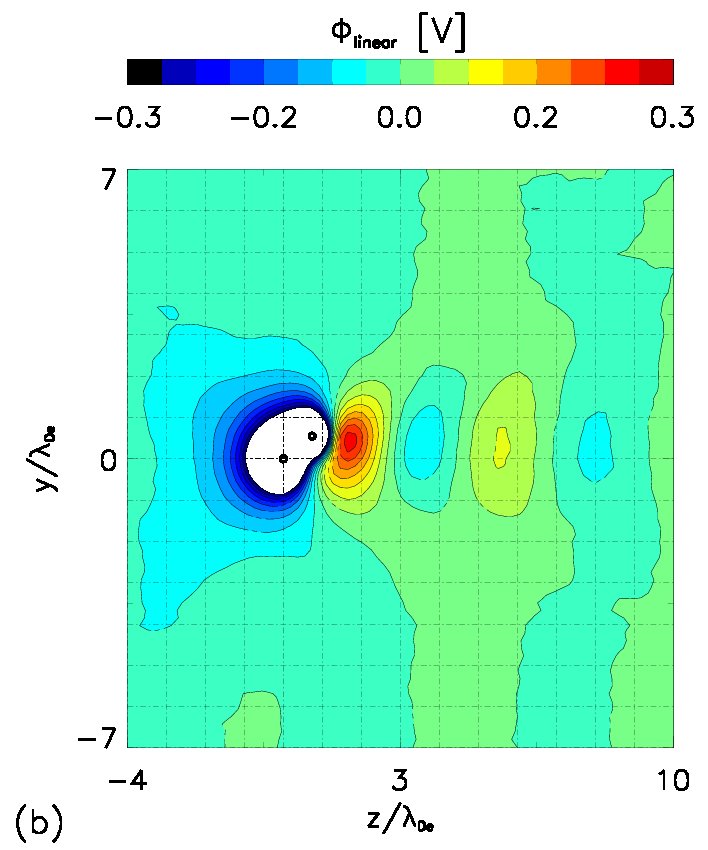}
\end{minipage}\hspace{0.018\textwidth}%
\begin{minipage}{0.33\textwidth}
\includegraphics[width=1.0\textwidth]{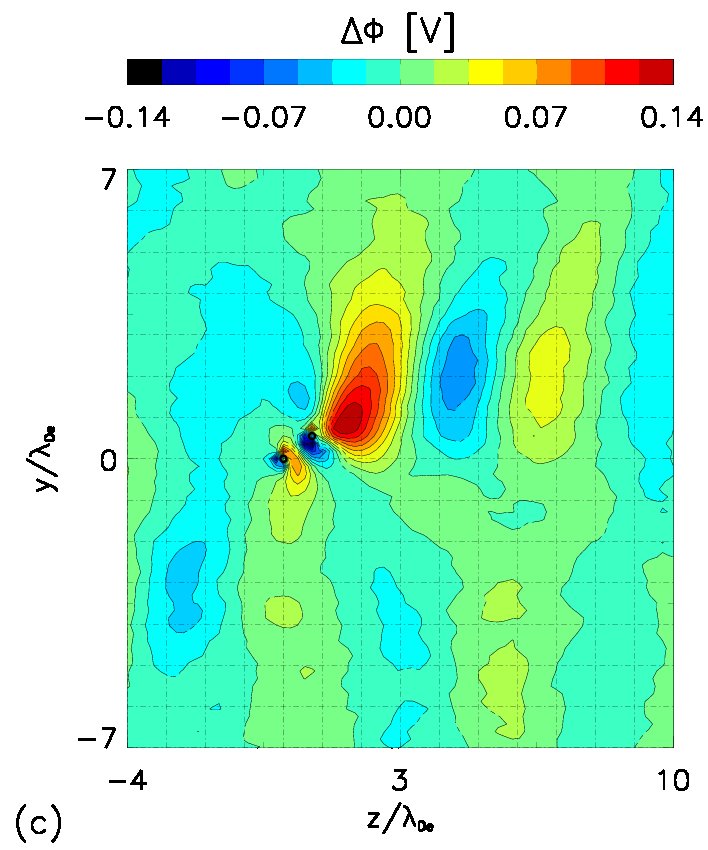}
\end{minipage} 
\begin{minipage}[b]{\textwidth}\caption{\label{fig:diff3} Same as Fig.~\ref{fig:diff}, but for subsonic flow $M=0.75$.}
\end{minipage}
\end{figure}

\begin{figure}[h]\hspace{-0.01\textwidth}%
\begin{minipage}{0.33\textwidth}
\includegraphics[width=1.0\textwidth]{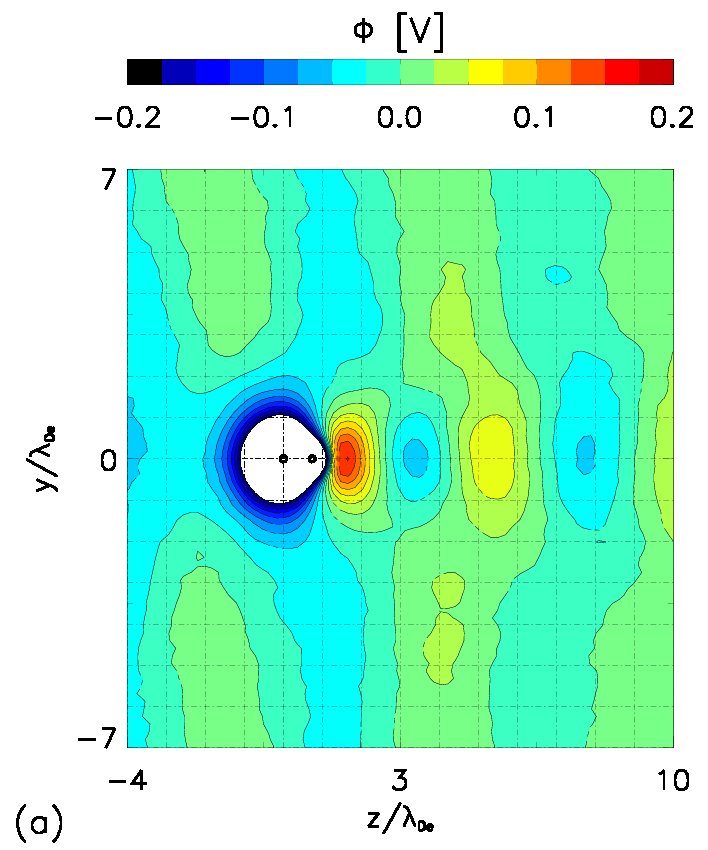}
\end{minipage}\hspace{0.018\textwidth}%
\begin{minipage}{0.33\textwidth}
\includegraphics[width=1.0\textwidth]{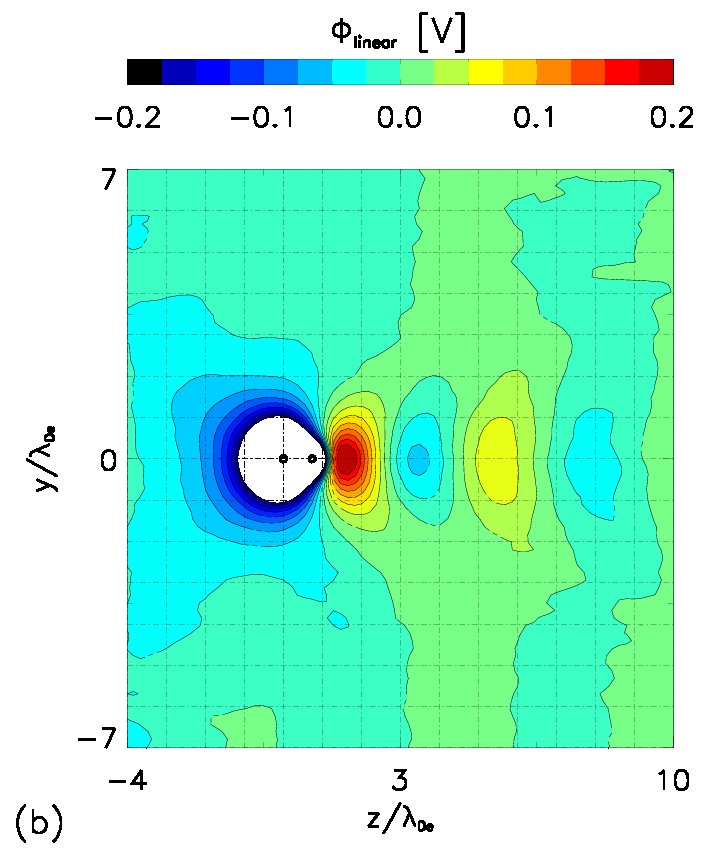}
\end{minipage}\hspace{0.018\textwidth}%
\begin{minipage}{0.33\textwidth}
\includegraphics[width=1.0\textwidth]{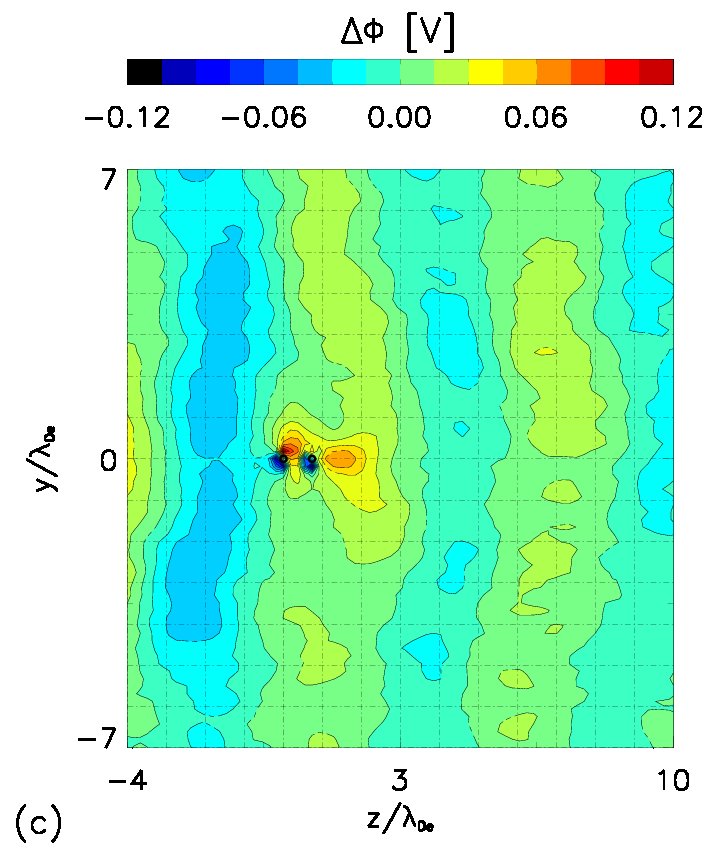}
\end{minipage} 
\begin{minipage}[b]{\textwidth}\caption{\label{fig:diff4}
Same as Fig.~\ref{fig:diff}, but for two flow-aligned grains and $M=0.75$.}
\end{minipage}
\end{figure}

\subsection{Consideration of shadowing effects} 
An important question for the OCP model is to determine to what extent the wake potential around two or more grains can be approximated by the linear superposition of single grain potentials. We will thus study the impact of non-linear effects in the multiple grain wakes by comparing  the results from PIC simulations of two grains with the results obtained by a linear combination of data from the corresponding simulation of a single grain.


In Fig.~\ref{fig:diff}(a), we show results from the PIC simulations of two grains for $T_e/T_i=30$ and $M=1.25$. The downstream grain is located off the symmetry axis at a distance $0.93 \lambda_{D_e}$ from the upstream grain. The charge of a downstream grain is only affected a little by the upstream grain, as the downstream grain is located out of the ion focus region \cite{miloch2010a}. 
In the linear combination of two wakes originating from a single grain calculation, the grain charges (and thus the grain potentials) are adjusted in such a way that they match the corresponding values from the simulations of two grains, see Fig.~\ref{fig:diff}(b). The constructed wake captures the general features of the results from the simulation of two grains, i.e., the position and approximate height of the potential maxima and minima. However, the small differences between the two results indicate that non-linear wake effects are present in the simulations of two grains.
In Fig.~\ref{fig:diff} (c) the potential difference $\Delta \Phi = \Phi_{\mathrm{linear}} - \Phi$ between the linear combination of wakes and the  original result is plotted.  In the system of two grains, the downstream grain influences the ion dynamics in the wake, and therefore leads to an asymmetric ion focus region, which can not be fully represented by a linear combination. The potential resulting from the linear combination overestimates the maximum in the wake by up to $20 \%$. Also the symmetry of the wake potential is broken.

The corresponding results for  two grains aligned with the flow that are separated by $0.74 \lambda_{D_e}$ are shown in Fig.~\ref{fig:diff2}. The charge on the downstream grain is strongly affected by ion focusing, and it is only about $30\%$ of the charge on the upstream grain. Thus in the linear combination of the wake, the contribution for the downstream grain has been adjusted accordingly. The constructed wake again represents well the topology of the wake from the simulations of two grains. In this case, the relative potential difference is up to $30 \%$ with the strongest discrepancies associated with the first maximum.

The agreement between constructed and simulated wake is worse for subsonic flows. Figs. \ref{fig:diff3} and \ref{fig:diff4} show corresponding results for the subsonic case $M=0.75$ for the downstream grain located off axis and aligned with the flow, respectively. In both cases, the differences $\Delta \Phi$ are up to $50 \% $ in the closest vicinity of the grains. 

In all considered cases the general features of the wake, such as positions of the maxima, are well reproduced. In the constructed wakes, the heights of the first maximum are overestimated, and for asymmetric grain arrangements, the symmetry of the wake further downstream from the grains is also modified. This is due to a single asymmetric ion focus region forming in a system of several grains with small separation which are not aligned with the flow.  Thus, constructing the wake from the linear combination of two wakes of single grains should be taken with care. To construct such a wake it is necessary to know \textit{a priori} the charge or potential on a downstream grain, which can be strongly influenced by the wake effects \cite{ikkurthi,miloch2010a}.  While the main features of the wake can be well represented by the linear combination of a single-grain wakes, there may be inaccuracy due to non-linear effects associated with ion focusing.

\subsection{Range of applicability of both simulation models}
In order to study the wake structure behind a grain in streaming plasmas, we have employed two complementary methods: (i) the  numerical evaluation of the linearized electrostatic potential that is created by a point-like charge, and (ii) self-consistent, full 3D particle-in-cell simulations. These methods have the best performance in different regimes. The LR method has the best convergence for small $T_r$ and large $\nu_{in}$ (and of minor importance small $M$), i.e., large Landau or collisional damping are optimal for the performance of the code. 
The PIC simulations are easiest for higher $M$  and small $T_r$, as the simulations for small $M$ and 
cold ions are often favorable condtitions for instabilities~\cite{Brackbill_1988}.

In the full PIC method, the ions have often reduced mass. The results with reduced ion mass are reliable, when the analysis is carried out in dimensionless units \cite{miloch2007}.
The PIC method, which includes self-consistent charging of the grain, accounts for a finite grain size, while the LR assumes a point-like dust grain with a given charge. The considered particle radius $a=0.185\lambda_{{D_e}}$ in the PIC simulations corresponds to the non-linear regime~\cite{hutch2011}.

The two approaches are complementary, and together allow for studies of a grain potential over a wide regime of parameters. Moreover, for single grain simulations, there is good agreement in the functional form of the wake (i.e., number and relative positions of the wake potential maxima and minima).
The main issue of the LR approach is the correct estimate of the grain charge as a function of $M$, and also for a grain that is shadowed by another grain located upstream.
A possible solution to overcome this issue could be the use of pre-computed tables obtained from self-consistent PIC simulations or making use of data from experiments~\cite{Carstensen}.
It is also worth mentioning that possible deviations from a shifted Maxwellian ion velocity distribution, as reported in~\cite{hutch2011,Schweig2001,ivlev2005}, need to be considered in developement of advanced models of the dust plasma interactions applicable for a wide range of plasma parameters.




\section{Summary and Outlook} 
In the present work, we have investigated the electrostatic potential distribution around a spherical dust grain in stationary and flowing plasmas by means of linear response calculations and particle-in-cell simulations. The streaming plasma leads to strong deviations from the statically screened Coulomb potential, giving rise to a distinct oscillatory wake structure behind the grain. 
For both sub- and supersonic flow velocities, positive ion wake charges downstream from the grain are present that can give rise to non-reciprocal dust grain interactions.
We have presented a systematic overview of the LR results for different flow velocities, electron-to-ion temperature ratios, and ion neutral collision frequencies. The LR results have been tested against self-consistent particle-in-cell simulations. The results from the PIC simulations for the wake pattern agree qualitatively with the LR data when the finite size of the grain is accounted for.

Downstream from the grain, a pronounced oscillatory wakefield with several maxima develops with increasing ion flow. Independent of the temperature ratio $T_e/T_i$, the peak positions are shifted linearly away from the grain when $M$ is increased.
Upstream and to the side from the grain, the potential decay can be approximated by a Yukawa potential. The effective screening length increases in both directions with the flow, but in contrast to earlier LR results~\cite{lampe2000}, it is found to be larger in the upstream direction than laterally from the grain. 

In a system of several grains in a plasma flow, the charge and potential on the downstream grains can be significantly modified by wake effects. The results from PIC simulations show that the wake behind the two grains can include non-linear effects, and that the charge  on the downstream grains can be significantly reduced.
While the general wake pattern can be represented by a linear combination of wakes behind a single grain to a rather good accuracy, local deviations up to $50\%$ of the potential are observed for subsonic flows in the closest vicinity of the grains. Thus, special care needs to be taken when applying LR results in OCP simulations, and at least the charge reduction on the downstream grains needs to be considered for the model to be credible.  

The influence of wake effects on collective many-particle behaviour in 3D plasma crystals has become a very timely question and is currently under ongoing experimental and theoretical investigation~\cite{kroll2010,schella2011,killer,ludwig2011}.
The presented wake potentials are currently employed in Dynamical Dust Simulations which accompany the experimental exploration of string formation and externally triggered phase transitions in 3D confined dusty plasma clouds~\cite{schella2011,killer}.
OCP simulations that include many particles require, however, an adequate adjustment of the charges of the individual grains, which are shadowed by the upstream grains. Such data may be obtained from the first principle PIC simulations or from experiments~\cite{Carstensen}. The development of such a grain-charging model is the subject of ongoing work.

Due to the high scalability of Coulomb systems, we expect these results to be of interest also to other types of multi-component plasmas. In particular, the generalization of the dynamically screened dust approach will open the way to a description of non-equilibrium quantum systems, such as warm dense matter, on the microscopic level. There, the classical dielectric function must be replaced by the corresponding quantum (Mermin) dielectric function~\cite{vlad11,ktmd,ludwig2010b}. This scheme then allows for the computation of collective many-body properties of strongly correlated ions embedded into a partially ionized quantum plasma of electrons by first principle molecular dynamics simulations.








\subsection*{Acknowledgment} 

The authors wish to thank I. H. Hutchinson for providing the data for Fig. \ref{fig:PICvsLR}. This work was supported by the Deutsche Forschungsgemeinschaft via grant LU 1586/1-1 and in part by the Norwegian Research Council, NFR, Grant No. 177570.

\end{document}